\documentclass[11pt, a4paper]{article}
\usepackage{amsmath}
\usepackage{graphicx}%,psfrag,epsf}
\usepackage{enumerate}
\usepackage{natbib}
\usepackage{amssymb}%
\usepackage{mathrsfs}
\usepackage{amsbsy}
\usepackage{amsthm}
\newtheorem{proposition}{Proposition}
\usepackage{enumitem}
\usepackage{booktabs}%% for tables
\usepackage{relsize}%% for fonts mathrm
\usepackage{graphicx}
 \usepackage{multirow}
\usepackage{rotating}
\usepackage{algorithm}
\usepackage{algorithmic}
\usepackage{fancyhdr}

\newcommand{\blind}{0}%

\newcommand{\ba}{{\mathbf{a}}}

\newcommand{\bb}{{\mathbf{b}}}
\newcommand{\be}{{\mathbf{e}}}
\newcommand{\bt}{{\mathbf{t}}}
\newcommand{\bp}{{\mathbf{p}}}

\newcommand{\bx}{{\mathbf{x}}}

\newcommand{\bz}{{\mathbf{z}}}

\newcommand{\bv}{{\mathbf{v}}}
\newcommand{\bX}{{\mathbf{X}}}

\newcommand{\bA}{{\mathbf{A}}}
\newcommand{\bB}{{\mathbf{B}}}
\newcommand{\bC}{{\mathbf{C}}}
\newcommand{\bD}{{\mathbf{D}}}

\newcommand{\bG}{{\mathbf{G}}}
\newcommand{\bT}{{\mathbf{T}}}

\newcommand{\bM}{{\mathbf{M}}}

\newcommand{\bZ}{{\mathbf{Z}}}
\newcommand{\bW}{{\mathbf{W}}}
\newcommand{\bR}{{\mathbf{R}}}

\newcommand{\bS}{{\mathbf{S}}}

\newcommand{\bI}{{\mathbf{I}}}
\newcommand{\bJ}{{\mathbf{J}}}
\newcommand{\bP}{{\mathbf{P}}}
\newcommand{\bQ}{{\mathbf{Q}}}

\newcommand{\un}{{uncorrelated}}

\newcommand{\uns}{{uncorrelatedness}}

\newcommand{\bta}{{\tilde{\mathbf{a}}}}

\newcommand{\bla}{{\boldsymbol{\lambda}}}

\newcommand{\trasp}{^{\prime}}
\newcommand{\eg}{{\emph{e.g.}}}
\newcommand{\vexp}{\text{Vexp}}

\DeclareMathOperator*{\argmin}{arg\,min}
\DeclareMathOperator*{\argmax}{arg\,max}

\newcommand{\Tr}{{\mathrm{tr}}}
\newcommand{\Lo}{{\text{L}_0}}
\newcommand{\Li}{\text{L}_1}
\newcommand{\Lt}{\text{L}_2}
\newcommand{\sq}{\frac{1}{2}}

\begin{document}
\def\spacingset#1{\renewcommand{\baselinestretch}%
{#1}\small\normalsize} \spacingset{1}
%%%%%%%%%%%%%%%%%%%%%%%%%%%%%%%%%%%%%%%%%%%%%%%%%%%%%%%%%%%%%%%%%%%%%%%%%%%%%%
\if0\blind
{
  \title{\bf Sparse Principal Component Analysis: a Least Squares approximation approach}
  \author{Giovanni Maria Merola\thanks{
    We would like to thank Dr. Alessio Farcomeni for making available his R code for the branch-and-bound search and Professor Bob Baulch of RMIT Vietnam for his useful comments.}\hspace{.2cm}\\
    Department of Economics, Finance and Marketing\\
    RMIT International University Vietnam\\
    702 Nguyen Van Linh\\
    Dist 7. Ho Chi Minh City, Vietnam\\
    (giovanni.merola@rmit.edu.vn)}
  \maketitle
} \fi

\if1\blind
{
  \bigskip
  \bigskip
  \bigskip
  \begin{center}
    {\LARGE\bf Sparse Principal Component Analysis: a Least Squares approximation approach}
\end{center}
  \medskip
} \fi

\bigskip
\begin{abstract}
Sparse Principal Components Analysis aims to find principal components with few non-zero loadings. We derive such sparse solutions by adding a genuine sparsity requirement to the original Principal Components Analysis (PCA) objective function. This approach differs from others because it preserves PCA's original optimality: \uns\ of the components and Least Squares approximation of the data. To identify the best subset of non-zero loadings we propose a Branch-and-Bound search and an iterative elimination algorithm. This last algorithm finds sparse solutions with large loadings and can be run without specifying the cardinality of the loadings and the number of components to compute in advance. We give thorough comparisons with the existing Sparse PCA methods and several examples on real datasets.
\end{abstract}

\noindent
{\it MSC-class:} {62H12; 62H25}
\newpage
\section{Introduction}
Principal Component Analysis (PCA) is one of the most frequently used methods for approximating a set of variables with few linear combinations of them, called \emph{principal components} (PCs) \citep[\eg][]{ize08}. PCA was originally introduced by \citet{pea} to find the "\emph{lines and planes of closest fit to systems of points}" and was later rediscovered by \citet{hot}, who wanted to find "\emph{some more fundamental set of independent variables {\rm[$\ldots$]} which determine the values the x's will take}." The PCs are estimated by minimising the sum of squares residuals of the approximation, in the words of \citet{pea} "\emph{a good fit will clearly be obtained if we make the sum of the squares of the perpendiculars from the system of points upon the line or planes a minimum}." Hence, PCA belongs to the class of multivariate methods that optimise the Least Squares (LS) criterion to find the estimates \citep{ten}.

The PCs are mutually uncorrelated and reproduce, or \emph{explain}, the most possible variance of the data. \citet{hot} showed that the PCs are also the components with orthonormal loadings that sequentially have largest variance. PCA has been popularised in this simpler form often without mention of the original LS fit requirement.

The weights of the PCs (\emph{loadings}) are used to interpret the PCs as meaningful characteristics of the data. For example, a component from a set of IQ test scores defined as $(\frac{1}{2}\; Inductive\ reasoning$ + $\frac{1}{2}\; Deductive\ reasoning)$ could be interpreted as the $Logic\ skills$ of a person. However, since  in most applications all the loadings are different from zero, convincing interpretations are difficult to find. The PCs would be easier to interpret if only few of the loadings were different from zero, that is, if they were \emph{sparse}.

The most common way to achieve sparseness is by \emph{thresholding} the loadings, that is, by discarding "small" ones (hereafter, for simplicity, we speak of the size of the loadings meaning the absolute value of the non-zero ones). Thresholding affects the optimality and \uns\ of the solutions in an unpredictable way \citep{cad}. Therefore, a number of \emph{Sparse Principal Component Analysis} (SPCA) methods for estimating components with few non-zero loadings have been proposed. The number of non-zero loadings is called \emph{cardinality} or $\Lo$ norm.

The existing SPCA methods maximise the variance of the components under cardinality constraints, in most cases also requiring that the loadings are orthonormal. However, because of the additional constraints, this is no longer equivalent to maximising the variance explained. So, the components obtained simply maximise a numerical property of the PCs that is irrelevant for summarising the information contained in the data \citep{ten}.

The lack of equivalence between objective function and variance explained created confusion in the SPCA literature on how components subsequent to the first one and the variance that they explain should be computed (\eg\ \citealp{mac} and \citealp{wan}). Furthermore, in the absence of specific constraints, the SPCA components are correlated. Correlated components are more difficult to interpret and the sum of the variance that they explain individually is larger then the variance that they explain together.

Finding the optimal SPCA solutions is a nonconvex NP-hard, hence intractable, problem \citep{mog}.
The existing SPCA methods use sophisticated numerical algorithms to approximately solve it. In this paper we will consider the estimation criteria used by these methods but not the numerical approximations proposed. A review of various SPCA methods can be found in \citet{tre13}.

In this paper we improve on the existing methods by deriving \un\ sparse PCs that minimise the LS criterion. In this method, to which we refer as LS-SPCA, the solutions are obtained by adding cardinality restrictions to the PCA problem. The solutions can be computed as a series of constrained Reduced Rank Regression components \citep[RRR,][]{ize}, for which we provide the closed form solutions.

Also for LS-SPCA the problem of finding the best set of loadings of a given cardinality is intractable. As a first approach, for its solution we suggest a simple greedy iterative Branch-and-Bound (BB) search based on that proposed by \citet{fra}. We will show that it can be applied to moderately large size problems in reasonable time.

SPCA should replace manual thresholding, which produces interpretable sparse solutions with few big loadings. This is not guaranteed by the constraints on the $\Lo$ norm and the sparse components computed may still be difficult to interpret.
For this reason, we propose an algorithm that computes the sparse solutions by iteratively eliminating the smallest loadings. This greedy algorithm, to which we refer to as Backward Elimination (BE), at each iteration eliminates the smallest loading until only loadings larger than a threshold remain. Since the size of the  loadings is not the only criterion to be considered, we also include rules for terminating the elimination when a maximum amount of variance explained is lost or a minimal cardinality is reached; these rules can be activated simultaneously.

The BE algorithm represents a departure from an optimal search. However, in terms of user's needs, the increased interpretability of the solutions should compensate the ensuing loss of variance explained. In the numerical section we show that the BE components compare well with the BB ones. Furthermore, with this flexible algorithm, cardinality, minimum size of the loadings and departure from the optimum can be controlled at the same time. These controls also eliminate the need for specifying in advance the cardinalities and the number of components to compute.

The selection of a sparse set of loadings is similar to model selection in Linear Regression, where there is a competition between parsimony of the model and variance of the response explained. In the same way, there exist similar sparse solutions that compete on variance explained and interpretability. The BE algorithm can be also used in an explanatory phase of the analysis  for comparing different solutions.
%Note that the first SPCA method proposed \citep[SCoT][]{\jol00} added the varimax criterion  used in a simplifying rotation algorithm.

The BB and BE algorithms are implemented in an R package, which will soon be released. In this implementation the sparse loadings can be restricted to include only a subset of variables and more than one loading can be eliminated when there are a large number of variables. The package also contains methods for producing summaries, plots and comparisons of the results. Some benchmark datasets are also included.

The original paper proposing LS-SPCA is still unpublished because reviewers seem to ignore that the original objective of PCA is the maximisation of the variance explained. Also journals editors seem to share the same lack of knowledge and to approve biased reviews, likely from authors of different SPCA methods. For example, Dr. Qiu, the chief editor of Technometrics, accepted a report which did not discuss at all the content but rejected the paper because I would ignore an article (by A. D'aspremont \emph{et al}.) which, instead, was cited in the references and the results of which were compared with mine (as in this version). He also accepted another report from a referee (who could barely write in English) who called the LS criterion "a new measure used ad-hoc". Dr Qiu rejected the paper on those grounds and added the reason that the algorithm is not scalable. Since computational efficiency is not in the scope of Technometrics, this also was unfair. The (kind) letter of complaint I sent to Dr. Qiu went unanswered. This is just to testify how frustrating publishing sometimes can be, thanks to unfair resistance from other fellow academics.

The rest of the paper is organized as follows: in the next section we formalise the SPCA problem into the LS optimisation framework and comparing it with the estimation criteria used in other SPCA methods. In Section \ref{sec:lsspca} we derive the closed form solutions for LS-SPCA method, also considering correlated ones. In Section \ref{sec:algos} we present the branch-and-bound search and the Backward Elimination algorithm. In Section \ref{sec:numcomp} we give extensive numerical comparisons with other SPCA methods on benchmark datasets and show the results of our methods on real life datasets of varying dimensions. Finally, in Section \ref{sec:concl} we give some concluding remarks.
\section{Sparse Principal Components Analysis}\label{sec:mod}
Statistical estimation theory requires that a measure of goodness of fit is defined and optimised. \citet{pea} explicitly adopts the Least Square criterion. Hotelling’s (1933) paper is quite intricate and the estimation
procedure used is not clearly stated. The solutions are the same as Pearson’s ones but are given as the components with orthonormal (that is with unit Euclidean norm (L2) and mutually orthogonal) loadings that have maximum variance. Due to its simplicity, this has become the standard definition with which PCA has been popularised among practitioners. It must be noted that Pearson’s solutions do not have the properties that he was looking for. In fact, in the introduction of his paper (page 417) he stated ”\emph{We shall consider only normally distributed systems of components having zero correlations and unit variances}”,
and the last two properties are paramount in his derivation. Obviously, the maximal variance of the components cannot be constantly equal to one and orthogonality of the loadings is not sufficient for \uns. Hence, the estimation did not solve the simplified problem.

In this section we first give an overview of the LS derivation of the PCs, because it is less known than others. Then we define the sparse PCA estimation criterion, or problem, by adding $\Lo$ constraints to the LS optimisation. Finally, we discuss the estimation criteria adopted by other SPCA methods.
\subsection{Notation}
In the following, we denote matrices with bold uppercase letters, $\bX$, and vectors with bold lowercase ones $\ba$. The columns of a matrix are denoted with the corresponding bold lowercase symbol, indexed accordingly, $\bx_j$. The symbol $\bI_k$ denotes the identity matrix of order $k$. The reference to the order is omitted when this is clear from the context.
$\Tr(\bS) = \sum_{j=1}^p s_{j,j}$ denotes the trace of a ($p\times p$) square matrix.
The $\Lt$ norm will be denoted as $||\cdot||$, so that $||\bX||= [\Tr (\bX\trasp\bX)]^{\frac{1}{2}}$. A "hat" on a symbol denotes an estimate and the subscript $[\cdot]$ denotes the rank of a matrix, so, for example, $\hat\bX_{[d]}$ is the estimate of a rank-$d$ approximation of the data matrix $\bX$. A "$^*$" denotes the globally optimal PCA estimates. The subscript $(d)$ denotes the first $(d-1)$ columns of a matrix, so that $\bA_{(j)}$ are the first $(j-1)$ columns of $\bA$.
\subsection{The PCA problem}
Given a matrix of $n$  observations on $p$ mean-centred variables $\bX$ ($n \times p$), PCA intends to find rank-$d$ ($d\leq p$) approximation of the data, $\hat\bX_{[d]}$. It is easy to prove that the approximation can be written as $\hat\bX_{[d]} = \bX\bA\bP\trasp$, where $\bA$ ($p\times d$) is the matrix of loadings,  $\bT = \bX\bA$ ($n\times d$) the matrix of the PCs and $\bP$ ($d\times p$) a matrix of coefficients.
The solutions are determined by minimising the LS criterion, that is, as:
\begin{equation*}
   \argmin\limits_{\mathrm{Rank}(\hat\bX_{[d]}) = d} ||\bX - \hat\bX_{[d]}||^2
    = \argmin\limits_{{\bA} \in \Re^{p\times d}} ||\bX - \bX\bA\bP\trasp||^2
\end{equation*}
The solutions are completely identified by the loadings because $\bP\trasp = \left(\bT\trasp\bT\right)^{-1}\bT\trasp\bX$ so that
the rank-$d$ approximation is equal to $\hat\bX_{[d]} = \bX\bA\left(\bA\trasp\bX\trasp\bX\bA\right)^{-1}\bA\trasp\bX\trasp\bX$.

The components can be constrained to be \un\ without loss of optimality.
In fact, by the principle of the \emph{extra sum of squares} a set of correlated components cannot explain more variance than the same number of uncorrelated ones (in the appendix we provide a proof of this well known result). Beside ease of interpretation, another advantage of \uns\ is that the resulting ordered PCs are the LS estimates for any number of components included in the model.

If we let $\bS = n^{-1}\bX\trasp\bX$ ($p\times p)$ denote the sample covariance matrix, under \uns\ constraints the PCA problem becomes:
\begin{eqnarray}\label{eq:optpca}
   &&\bA = \argmin\limits_{\mathrm{Rank}(\hat\bX_{[d]}) = d} ||\bX - \hat\bX_{[d]}||^2 =    \argmax\limits_{{\bA} \in \Re^{p\times d}}||\hat\bX_{[d]}||^2  =
    \argmax\limits_{{\ba_j} \in \Re^{p}} \sum_{j = 1}^d
    \frac{\ba\trasp_j\bS \bS\ba_j}{\ba\trasp_j\bS\ba_j}\\
    &&\text{subject to}\; \ba\trasp_j\bS\ba_k=0,\; j \neq k,\nonumber
\end{eqnarray}
where the summation in the last term derives from the uncorrelatedness of the components. It follows that the total variance explained can be broken down into the sum of the variances explained by each component. Therefore the LS criterion is equivalent to the maximisation of the variances explained by each component, $\vexp(\bt_j)$. Hence, the PCA problem can be equivalently written as:
%$\frac{\ba\trasp_j\bS \bS\ba_j}{\ba\trasp_j\bS\ba_j}$.
\begin{eqnarray}\label{eq:vexp}
   &&\ba_j =  \argmax\limits_{{\ba_j} \in \Re^{p}} \vexp(\bt_j) = \argmax\limits_{{\ba_j} \in \Re^{p}} \frac{\ba\trasp_j\bS \bS\ba_j}{\ba\trasp_j\bS\ba_j}.\\
    &&\text{subject to}\; \bb_j\bS\trasp\ba_k = \delta_{jk}\,\nonumber
\end{eqnarray}
\subsection{The PCA solutions}
If no other constraints are added to Problem (\ref{eq:optpca}), the PCA loadings are proportional to the eigenvectors of $\bS$, $\{\bv_j, j=1,\ldots,d\}$, corresponding to the $d$ largest eigenvalues taken in non increasing order, $\{\lambda_1 \geq \lambda_2\geq \ldots\geq \lambda_d\}$. Then it follows that the loadings are orthonormal and the variance explained by each PC is equal to the corresponding eigenvalue, because Equation (\ref{eq:vexp}) simplifies to:
\begin{equation*}
\vexp^*(\bt_j) = \frac{\bv\trasp_j\bS \bS\bv_j}{\bv\trasp_j\bS\bv_j} = \frac{\bv\trasp_j\bS\bv_j}{\bv\trasp_j\bv_j} = \lambda_j.
\label{eq:vexpaseigvals}
\end{equation*}
Hence the variance explained by a PC is equal to its variance. This property has led to the popularisation of PCA simply as the method that finds the components with orthonormal coefficients that have sequentially maximal variance. Consequently, the PCA problem is often defined as:
\begin{eqnarray}\label{eq:pcahot}
    &&\ba_j = \argmax\limits_{{\ba_j} \in \Re^{p}} \ba_j\trasp\bS\ba_j, j = 1,\ldots,d\\
    &&\text{subject to}\; \ba_j\trasp\ba_k = \delta_{jk},\nonumber
\end{eqnarray}
where $\delta_{jk}$ is the Kronecker delta. \citet[][page 87]{ten} warns about this formulation of the PCA problem by stating: "\emph{Nevertheless, it is undesirable to maximize the variance of the components rather than the variance explained by the components, because only the latter is relevant for the purpose of finding components that summarize the information contained in the variables.}"
\subsection{The Least Squares Sparse PCA problem}
The LS-SPCA Problem is obtained by constraining the cardinality of the loadings in Problem (\ref{eq:optpca}), which gives:
\begin{eqnarray}\label{eq:lsspca}
    &&\bA = \argmin\limits_{{\bA} \in \Re^{p\times d}} ||\bX - \bX\bA\bP\trasp||^2
    = \argmax\limits_{{\bA} \in \Re^{p\times d}}\sum_{j=1}^d
    \frac{\ba\trasp_j\bS \bS\ba_j}{\ba\trasp_j\bS\ba_j} =
    \argmax\limits_{{\bA} \in \Re^{p\times d}}\sum_{j=1}^d\vexp(\bt_j) \\
    &&\text{subject to}\; \Lo(\ba_j) \leq c_j\ \text{and}\ \ba_j\trasp\bS\ba_k=0,\ j\neq k,\nonumber
\end{eqnarray}
where $c_j < p$ are the maximal cardinalities allowed.

As we will show in Section \ref{sec:lsspca}, under the cardinality constraints the loadings $\vexp(\bt_j)$ no longer simplifies to $(\ba\trasp_j\bS\ba_j)/\ba\trasp_j\ba_j$ and the solutions must be obtained by maximising \emph{Vexp} in Equation (\ref{eq:vexp}) directly.
\subsection{Other SPCA problems}\label{sec:spcaold}
In the existing SPCA methods the components are computed by maximising the variance of the components under cardinality constraints. Hence, the sparse PCA problem is defined by adding cardinality constraints to the simplified PCA problem (\ref{eq:pcahot}), which gives:
\begin{eqnarray}\label{eq:spca}
    &&\bb_j = \argmax\limits_{{\bb_j} \in \Re^{p}} \bb_j\trasp\bS\bb_j, j = 1,\ldots,d\\
    &&\text{subject to}\; \bb_j\trasp\bb_k = \delta_{jk}\, \text{and}\, \Lo(\bb_j) \leq c_j ,\nonumber
\end{eqnarray}
for some parameters $c_j$ ($c_j < p$). Clearly, this problem is not analogous to Problem (\ref{eq:lsspca}) because it implicitly assumes that the solutions are eigenvectors of $\bS$, which they cannot be under sparsity requirements.

The SPCA solutions of a given cardinality $c_j$ are the first eigenvectors of the
($c_j \times c_j$) principal submatrix of $\bS$ with the largest maximal eigenvalue, subject to the constraints \citep[][Proposition 1]{mog}. Hence the SPCA probelm can also be written in terms of the non-zero loadings $\tilde\bb_j$ as
\begin{eqnarray*}\label{eq:spcasol}
    &&\tilde\bb_j = \argmax\limits_{{\tilde\bb_j} \in \Re^{c_j}}
    \tilde\bb_j\trasp\bD_j\tilde\bb_j,\; j = 1,\ldots,d\\
    &&\text{subject to}\; \tilde\bb_j\trasp\tilde\bb_k = \delta_{jk},\nonumber
\end{eqnarray*}
where $\bD_j$ is the principal submatrix of $\bS$ corresponding to indices of the sparse loadings.
Hence, the SPCA problem boils down to finding the sets of indices, $ind_j$, that give the principal submatrix of $\bS$ with largest maximum eigenvalue.

In some SPCA methods \citep[\eg][]{mog} the orthogonality constraints are omitted from the problem. In this case, trivial solutions are avoided by computing the solutions subsequent to the first one on the covariance matrix \emph{deflated} in different ways. In general, there is no agreement on which is the correct deflation to use (see \citealp{mac}, and \citealp{wan}, for a discussion).
It should be noted that, the orthogonality constraints ensure that
a full set of sparse components explains all of the data variance, while this is not guaranteed if the components are not \un\ and the loadings not orthogonal. We observe that a necessary condition for orthogonality is that the cardinality of the loadings in not smaller than their rank. This condition, as we will show in the next section, applies also to the \uns\ constraints. The solutions to Problem (\ref{eq:spca}) subsequent to the first one are given by
\begin{eqnarray*}
    &&\bb_j = \argmax\limits_{{\bb_j} \in \Re^{p}}
    \bb_j\trasp\left(\bI - \bB_{(j)}(\bB_{(j)}\trasp\bB_{(j)})^{-1}\bB_{(j)}\trasp\right) \bS
\left(\bI - \bB_{(j)}(\bB_{(j)}\trasp\bB_{(j)})^{-1}\bB_{(j)}\trasp\right) \bb_j\\
    &&\text{subject to}\; \bb_j\trasp\bb_j = 1\, \text{and}\, \Lo(\bb_j) \leq c_j. \nonumber
\end{eqnarray*}
These solutions correspond to the \emph{additional variance} deflation, derived by \citet{mac} from different properties.

Some authors assume without justification that the variance of the sparse components is equal to the variance explained (\eg\ \citealp{dasp08}). In other cases the maximisation of the variance of the components is derived from different problems. \citet{zou} adopt a LS optimisation subject to $\Li$ constraints, in a Lasso fashion.
However, they also constrain the coefficients $\bp_j$ to have unit variance, that is, they require that $\|\bp_j\|^2 =1$. Since $\bp_j = \bS\bb_j/(\bb_j\trasp\bS\bb_j)$, these constraints are equivalent to requiring that
\begin{equation*}
\frac{\bb_j\trasp\bS\bS\bb_j}{\bb_j\trasp\bS\bb_j} = \frac{\bb_j\trasp\bS\bb_j}{\bb_j\trasp\bb_j},
\end{equation*}
where the right hand side is equivalent to the variance of the components with normal loadings. In this way, the scope of the objective is limited because, by the Cauchy-Schwartz inequality:
\begin{equation*}
\frac{\bb_j\trasp\bS\bS\bb_j}{\bb_j\trasp\bS\bb_j} \geq \frac{\bb_j\trasp\bS\bb_j}{\bb_j\trasp\bb_j},
\end{equation*}
for any square matrix $\bS$. It should be noted that \cite{zou} want to achieve Lasso type "shrinkage" estimates and not traditional LS ones.

\citet{she} justify the maximisation of the variance of the components considering that the PCA solutions can be derived from the maximisation of the trace of the approximation of the covariance matrix. The approximation is $\hat\bS^*_{[d]} = {\hat\bX^{*\prime}}_{[d]}\hat\bX^*_{[d]} = \sum_{j=1}^d \bv_j\bv_j\trasp\lambda_j$, with $\lambda_j = \bv_j\trasp\bS\bv_j$ and $\bv_j\trasp\bv_k=\delta_{jk}$.
Therefore, they seek the orthonormal sparse vectors $\bb_j$ with maximal variance so that the approximate covariance matrix is $\bS_{[d]}= \sum_{j=1}^d \bb_j\bb_j\trasp(\bb_j\trasp\bS\bb_j)$. However, also in this case, the LS-SPCA loadings, $\ba_j$ dominate the SPCA solutions. In fact, if we define the orthonormal vectors $\be_j = \bS^\sq\ba_j/(\ba_j\trasp\bS\ba_j)^\sq$, we can write the LS-SPCA approximation as
$\hat\bS_{[d]} = \sum_{j=1}^d \be_j\be_j\trasp \rho_j$ with $\rho_j =  \be_j\trasp\bS\be_j = \ba_j\trasp\bS\bS\ba_j/(\ba_j\bS\ba_j)$. Then,
\begin{equation*}
\Tr\left(\hat\bS_{[d]}\right) = \sum_{j=1}^d \frac{\ba_j\trasp\bS\bS\ba_j}{\ba_j\bS\ba_j} \geq \sum_{j=1}^d \frac{\bb_j\trasp\bS\bS\bb_j}{\bb_j\bS\bb_j}
\geq \sum_{j=1}^d \frac{\bb_j\trasp\bS\bb_j}{\bb_j\bb_j}= \Tr\left(\bS_{[d]}\right),
\end{equation*}
where  $\ba_j$ and $\bb_j$ have the same cardinality. Even not considering this result, it is still unexplained in which sense the resulting vectors $\bb_j$ could be used as loadings for the components, because in this case the approximation of the variance would be $\bS_{[d]} = \bS\bB(\bB\bS\bB)^{-1}\bB\bS$ and not the one optimised. Furthermore, when the components are correlated the trace of the approximation is no longer the sum  of the individual approximations but the sum of the diagonal elements of the  matrix $\bS_{[d]}$. 
\section{Least Squares Sparse PCA solutions}\label{sec:lsspca}
If we assume that the indices of the sparse loadings of the $d$ required components, $ind_j$,  are known, the sparse components will be combinations of only the corresponding variables, denoted with the matrices $\bW_j\ (n\times c_j)$. Then, the sparse components can be written as $\bt_j = \bW_j\bta_j$, where the $\bta_j$ are the vectors of dimension $c_j$ containing only the non-zero loadings. With this notation we can write the individual LS-SPCA problems as:
\begin{align}\label{eq:lsspcaind}
        &\bta_j = \argmin\limits_{{\bta_j} \in \Re^{c_j}}
        ||\bX - \bW_j\bta_j\bp\trasp_j||,
\; j = 1,\ldots, d\\
&\text{subject to}\; \bT_{(j)}\trasp\bW_j\bta_j = \mathbf{0},\;  j > 1,
\end{align}
where $\bT_{(j)}$ is the matrix containing the first $(j-1)$ components ($\bT_{(1)}=\mathbf{0}$). Let $\bJ_j$ ($p\times c_j$) be the matrices formed by the columns of the $p$ dimensional identity matrix with indices in $ind_j$, then we can write $\bW_j = \bX \bJ_j$ and the full sparse loadings as $\ba_j = \bJ_j\bta_j$. With this notation, the SPCA problem defined in Equation (\ref{eq:lsspcaind}) can be written as
\begin{align}\label{eq:uspcals}
&\argmin\limits_{{\bta_j} \in \Re^{c_j}}
        ||\bX - \bW_j\bta_j\bp\trasp_j|| = \argmax\limits_{{\bta_j} \in \Re^{c_j}} \frac{\bta_j\trasp\bJ_j\trasp\bS\bS\bJ_j\bta_j} {\bta_j\trasp\bJ_j\trasp\bS\bJ_j\bta_j}
        = \argmax\limits_{\bta_j \in \Re^{c_j}\; \ba_j=\bJ_j\bta_j} \frac{\ba_j\trasp\bS\bS\ba_j}{\ba_j\trasp\bS\ba_j},
        \; j = 1,\ldots, d\\
&\text{subject to}\ \bR_{j}\ba_j = \mathbf{0},\text{for}\ j > 1,\nonumber
\end{align}
where $\bR_j = \bA_{(j)}\trasp\bS\bJ_j$ defines the \uns\ constraints, with $\bA_{(j)}$ being the first $(j-1)$ loadings. Hence the sparse PCs maximise \vexp\ defined in Equation (\ref{eq:vexp}), under the constraints.

Problem (\ref{eq:uspcals}) can be seen as a series of constrained rank-one Reduced-Rank Regression problems where the regressors are the columns of the $\bW_j$ matrices. It is well known that the first solution is the eigenvector $\bta_1$ satisfying:
\begin{equation}\label{eq:uspcalssol1}
(\bW\trasp_1\bW_1)^{-1}\bW\trasp_1\bX\bX\trasp\bW_1\bta_1 = \phi_{\text{max}} \bta_1,
\end{equation}
where $\phi_{\text{max}}$ is the largest eigenvalue. This solution is unique as long as the variables in $\bW_1$ are not multicollinear. Hereafter we exclude the possibility that a matrix $\bW_j$ is not full column rank because that set of variables should be discarded and a full rank one sought.
The sparse loadings can be computed also if only the covariance matrix $\bS$ is known. In fact, Equation (\ref{eq:uspcalssol1}) can be written as:
\begin{equation}\label{eq:uspcalssol2}
\bD_1^{-1}\bJ_1\trasp\bS\bS\bJ_1\bta_1 = \phi_{\text{max}} \bta_1,
\end{equation}
where $\bD_j = \bW_j\trasp\bW_j = \bJ\trasp_j\bS\bJ_j$ is the covariance matrix of the variables with index in $ind_j$, which is invertible for the full rank assumptions.

The following solutions can be found by applying the \uns\ constraints to the RRR Problems (\ref{eq:uspcals}) as in constrained multiple regression
(\eg, see \citealp{rao}, or \citealp{mag}, Th. 13.5, for a more rigorous proof).
%\citep[\eg, see][Th. 13.5, for a more rigorous treatment]{rao,mag}.
In the appendix we show that these are given by the eigenvectors satisfying
\begin{equation}\label{eq:optspca}%\bB_j
%\left[\bI_{c_j} - \bD_j^{-1} \bR\trasp_j(\bR_j\bD_j^{-1}\bR\trasp_j)^+\bR_j\right]
\bC_j\bD_j^{-1}\bJ\trasp_j\bS\bS\bJ_j\bta_j = \phi_{\text{max}}\bta_j,
\end{equation}
where $\bC_j = \bI_{c_j} -  \bD_j^{-1} \bR\trasp_j(\bR_j\bD_j^{-1}\bR\trasp_j)^+\bR_j$, with $\bC_1 = \bI_{c_1}$, and the subscript "+" denotes a generalized inverse.
The solutions exist because $\bR_j$ spans the space of $\bW_j$. In this derivation we assume that $\bR\trasp_j\bR_j$ is singular, otherwise $\bR_j\bta_j=\mathbf{0}$ can never be satisfied. This means that \uns\ can only be achieved if the cardinalities satisfy $c_j \geq j$. The LS-SPCA solutions can be computed from the leftmost eigenvector, $\bb_j$, of the symmetric matrices
$(\bC_j\bD_j^{-1})^{\frac{1}{2}}\bJ\trasp_j\bS\bS\bJ_j(\bC_j\bD_j^{-1})^{\frac{1}{2}}$%{\frac{1}{2}}$
%$(\bJ\trasp_j\bS\bS\bJ_j)^{\frac{1}{2}} \bC_j\bD_j^{-1} (\bJ\trasp_j\bS\bS\bJ_j)^{\frac{1}{2}}$
%$(\bJ\trasp_j\bS\bS\bJ_j)^{\frac{1}{2}} \bD_j^{-1/2} \bC_j\bD_j^{-1/2} (\bJ\trasp_j\bS\bS\bJ_j)^{\frac{1}{2}}$
%as $\bta_j = (\bJ\trasp_j\bS\bS\bJ_j)^{\frac{1}{2}} \bb_j$.
as $\bta_j = (\bC_j\bD_j^{-1})^{\frac{1}{2}} \bb_j$.

The above derivation shows that the sparse components that explain the most variance are not eigenvectors of submatrices of the covariance matrix and that their variance is no longer equal to the variance that they explain.

As mentioned above, in LS-SPCA the \uns\ constraints require that the cardinality of the components is not less than their order. Correlated component of lower cardinality can be computed by
applying LS-SPCA to the residual of $\bX$ orthogonal to the previous components, $\bX_j = \bI - \bT_{(j)}(\bT_{(j)}\trasp\bT_{(j)})^{-1}\bT_{(j)}\trasp\bX$. While correctly used in iterative PCA algorithms, such as Power Method and NIPALS \citep{wol}, for example, under sparsity constraints this approach does not  maximize \vexp\ in Equation (\ref{eq:vexp}), but the approximations
\begin{equation}\label{eq:vexpno}
    \frac{\ba\trasp_j\bS_j \bS_j\ba_j}{\ba\trasp_j\bS\ba_j},
\end{equation}
hence the suboptimality.

In the appendix we show that these correlated components are given by the eigenvectors satisfying
\begin{equation}\label{eq:cspcalssol}
\bD_j^{-1}\bJ\trasp_j\bS_j\bS_j\bta_j = \phi_{\text{max}} \bta_j,
\end{equation}
where the $\bS_j$ are the covariance matrices computed from the residuals $\bX_j$. When needed, we will refer to these solutions as Least Squares Correlated Sparse Principal Components Analysis (LS-CSPCA).

Following the same approach, correlated sparse components with orthogonal loadings can also be found. In the appendix we show that these solutions are the eigenvectors satisfying:
\begin{equation*}\label{eq:ospcasol}
    \bD_j^{-1}\bJ_j\trasp\left[\bI_{p} - \bA_{(j)}
    \left(\bA_{(j)}\trasp \bJ_j\bD_j^{-1}\bJ_j\trasp\bA_{(j)}\right)^{-1}
    \bA_{(j)}\trasp\bJ_j\bD_j^{-1}\bJ_j\trasp \right]
    \bS_j\bS_j\bJ_j\bta_j = \phi_{max}\bta_j.
\end{equation*} \section{Determining the indices of the sparse loadings}\label{sec:algos}
Solving the LS-SPCA problem requires determining the optimal set of indices for $d$ components. The cardinality constraints make it a non-convex problem, which cannot be efficiently solved. We consider first a greedy Branch-and-Bound search (BB) then we consider a greedy Backward Elimination algorithm designed to replace manual thresholding.
\subsection{LS-SPCA(BB): a branch-and-bound search for the optimal loadings}
If the cardinalities of each component are specified, locally optimal subsets of loadings can be found through a greedy BB search based on that proposed by \citet{fra} using \vexp\ in Equations (\ref{eq:vexp}) as bounding function. At each node a subsets of variables of cardinality greater than the required one is discarded if the solution explains less variance than the current upper bound. The search continues on to subsets of smaller size until the best set of the required cardinality is found. This procedure leads to the optimal solution for each component because eliminating a variable from a regression model cannot yield a higher regression sum of squares. For correlated components, we use the variance explained by the approximate solutions, Equation (\ref{eq:vexpno}), because the true variance explained, Equation (\ref{eq:vexp}), is not monotonically larger.

The search can start from a subset of indices instead of the complete one, if the analyst has a tentative solution in mind. The algorithm can be speeded up by sorting the variables with respect to the variance of the residuals that they individually explain. Note that the solutions are only locally optimal for each component, because the search does not explore all combinations of loadings for the given number of components. We will refer to the solutions obtained using this BB algorithm as LS-SPCA(BB) and LS-CSPCA(BB) for the uncorrelated and correlated solutions, respectively.

LS-SPCA(BB) is not computationally efficient and can only be run on medium or small size problems within reasonable time. \citet{fra} gives an account of the computational times taken by the BB search for his method; in LS-SPCA(BB) it would take longer because the solutions are more complex to compute. In spite of this, the BB can be run on moderate size problems which are typically the ones in which the loadings are interpreted.
\subsection{LS-SPCA(BE): a backward elimination for thresholding the loadings}
%Like other SPCA methods, LS-SPCA(BB) requires  Also, in most cases the best cardinalities are not known in advance and a number of solutions of different cardinalities must be computed and examined to find the ones that best suit the analysis \citep[see][for some discussion on this topic]{fra}. Furthermore,
Like all other SPCA methods, LS-SPCA(BB) has the drawback that some of the sparse loadings computed may be small, making the solutions still difficult to interpret. As a matter of fact, SPCA should replace thresholding, which gives only big sparse loadings. Hence, we consider the problem of determining the sparse solutions so that the sparse loadings are larger than a given threshold value. Let the thresholds for each dimension be denoted as $\tau_j$ ($0 < \tau_j \leq 1$), then the problem can be formalized as
\begin{align}\label{eq:tspca}
    &\ba_j = \argmin\limits_{{\ba_j} \in \Re^{p}}
    ||\bX - \bX\ba_j\bp\trasp_j||,\; j=1,\ldots, d \\
    &\text{subject to}\;
    \left\{
        a_{ij} = 0\; \text{or}\; \frac{|a_{ij}|}{{\rm L}_{\cdot}(\ba_j)} > \tau_j
    \right\}\;
    \text{and}\, \bt\trasp_j\bt_k = 0\; j \neq k.
\nonumber%\sum_{i\not\in ind_j}|a_{ij}| = 0,
\end{align}
where ${\rm L}_{\cdot}(\ba_j)$ denotes a norm, typically the $\Li$ or $\Lt$ norm.
This problem is NP-hard and BB type searches are difficult to derive because bounding functions are not obvious. However, if the variables are standardized to the same length, it can be expected that eliminating a small loading will not decrease Vexp by much. Therefore, we suggest a simple greedy backward elimination algorithm to iteratively eliminate the smallest sparse loadings from a solution until only ones larger than a given threshold are left. In this procedure it is not necessary to specify the cardinality of the solutions in advance and the the elimination can be stopped by criteria different from the minimum threshold. For example, if only $j$ non-zero loadings are left, the elimination must be stopped to maintain \uns.

\citet{wan} proposed a backward elimination algorithm for SPCA based on different criteria. They mention examples in which eliminating small loadings is highly unreliable for SPCA. We believe that this is true when the components are correlated and the loadings are computed as pseudo-eigenvectors of a deflated matrix. In our studies we found our BE procedure very reliable and, in some cases, better than the BB search, as will be shown in the examples in Section \ref{sec:numcomp}.

The BE algorithm is outlined in Algorithm \ref{algo:spcabe}.
\algsetup{indent=2em}
\begin{algorithm}[H]\caption{LS-SPCA(BE)}
    \begin{algorithmic}
    \STATE {\bf initialize}
    \STATE {\hspace{1em}\bf Stopping rules for the number of components}
	\STATE{\hspace{2em}$\quad d$ }\COMMENT{the number of components to compute}
    \STATE{\hspace{2em}$\quad mv$ }\COMMENT{optional, minimum variance cumulated explained for ending the algorithm}
    \STATE {\hspace{1em}\bf Stopping rules for elimination. Can be different for each component}
	\STATE{\hspace{2em}$\quad ind_j$ }\COMMENT{starting set of indices}
	\STATE{\hspace{2em}$\quad \tau_j$ } \COMMENT{minimum absolute value of the sparse loadings}
	\STATE{\hspace{2em}$\quad k_j \geq 1$ } \COMMENT{minimum cardinality of the sparse loadings}
	\STATE{\hspace{2em}$\quad mvl_j$ } \COMMENT{optional maximum relative loss of variance explained}
    \STATE {\bf end initialize}
    \FOR{$j = 1\ \TO\ d$}
        \STATE{Compute $\ba_j$ as the j-th LS-SPCA solution for $ind_j$}
        \STATE{$\text{Vexpfull}_j = \vexp(\ba_j)$}
        \WHILE{ $\min_{i \in ind_j} |a_{ij}| > \tau_j$ \AND $length(ind_j) > k_j$}
            \STATE{$indold_j = ind_j$, ${\bf aold}_j = \ba_j$ }
            \STATE{$k$:\: $|a_{kj}| \leq |a_{ij}|, i \in ind_j$}
            \STATE{$ind_j = ind_j\backslash k$}
            \STATE{Compute $\ba_j$ as the j-th LS-SPCA solution for $ind_j$}
            \IF{$1 - \vexp(\ba_j)/\text{Vexpfull}_j > mvl_j$}
                \STATE{$ind_j = indold_j$, $\ba_j = {\bf aold}_j$}
                \STATE{{\bf break}}
            \ENDIF
        \ENDWHILE
        \IF{$\sum_{i=1}^j \vexp(\ba_i)  \geq \text{mvl}$}
            \STATE{$d = j$}
            \STATE{{\bf break}}
        \ENDIF
    \ENDFOR
    \end{algorithmic}\label{algo:spcabe}
\end{algorithm}
Depending on the choice of the thresholds, trimming may cause too large a loss of Vexp or yield solutions with too few non-zero loadings. Therefore, we include in the algorithm two optional additional stopping rules for trimming: one based on the required minimum cardinality (which is needed if \un\ components are sought) and the other based on the maximal loss of Vexp induced by the last trimming. These stopping criteria can be used instead of specifying the thresholds or additionally to it.
Specifying in advance the number of components to compute can be difficult. So, we include also an optional rule for stopping the algorithm when a specified proportion of total variance explained is reached. This rule can be used together with a specified maximum number of components to be computed.
The basic LS-SPCA Backward Elimination algorithm (LS-SPCA(BE)) is outlined in Algorithm \ref{algo:spcabe}:

Hence, the LS-SPCA(BE) algorithm can be used as a flexible tool without the need of specifying in advance the number of components to compute and their cardinality. In order to decrease the computational time for large matrices, more than one loading can be trimmed at each iteration. In this case, better results can be reached if, instead of terminating trimming when the cardinality is smaller than the number of loadings to be trimmed, the process is continued by trimming the remaining loadings individually until a stop rule is reached. Trimming can also be started from a subset of indices, if there is a reason for excluding some of the variables from a component.

For \un\ components, the minimal cardinalities must be set so as $k_j\ge j$, otherwise \uns\ cannot be achieved (unless the reduced set is multicollinear). In this case, solutions fully trimmed to a given threshold  may not be obtained. The stoprule for ending trimming can be defined with respect to different criteria, for example the loss of Vexp from the initial solution (as in Algorithm \ref{algo:spcabe}) or the loss of Cumulative Vexp from the corresponding PCA value. There is no obvious rule for choosing the thresholds $\tau_j$. However, if the loadings are computed to have unitary $\Lt$ norm, setting $\tau_j > 1/\sqrt{c}$ will ensure a cardinality lower than $c$ (almost surely).  In some cases, the size of the loadings is easier to evaluate if they are expressed as percentage \emph{contributions}, that is, are standardized to unit $\Li$ norm. In this case, setting $\tau_j > 1/c$ will ensure a cardinality lower than $c$. For this reason, the choice of the minimum cardinality and of the threshold must be considered together and later components require a lower threshold than the first ones. The minimal total variance to be explained for ending the algorithm can be chosen with respect to the Vexp explained by the PCs. Note that trimming is designed for components computed from correlation matrices. If a covariance matrix is used, different thresholds for every variable should be used. 
\section{Numerical comparisons and examples}\label{sec:numcomp}
In this section we first compare LS-SPCA with other existing SPCA methods on two benchmark datasets. Then we show the results of LS-SPCA on several publicly available datasets. All the examples were computed on correlation matrices but we refer to the variance explained rather than to the correlation explained for uniformity with the literature. In order to make fair comparisons, the variance explained was recomputed according to Equation (\ref{eq:vexp}) for all the solutions from other methods. Loadings smaller than 0.001 in absolute value are not shown.
\subsection{Comparison with other methods}%
In this section we compare the LS-SPCA results with those of other SPCA methods published for the only two benchmark datasets available in the literature, the famous Pitprop dataset \citep{thu} and an artificial one proposed by \cite{zou}. Therefore, the list of methods included is necessarily not exhaustive but it contains all the methods which, to our knowledge, were tested on these datasets. Unfortunately, the datasets are small in size, therefore differences in the variance explained are also small. Furthermore, the solutions in most cases were obtained for different cardinalities. This means that we can only compare LS-SPCA with each one of them and comparisons between different SPCA methods are not possible.

In our comparison we include some methods that relax the cardinality constraints by replacing it with constraints in the  $\Li$ norm of the loadings. These are: SCoTLAS \citep{jol03,tre}; \emph{SPCA} \citep{zou} who used a Lasso approach, and \citet{dasp}, DSPCA. Other methods that tackle the $\Lo$ constraints are: \citet{mog}'s greedy and exact Branch-and-Bound algorithms, GSPCA and ESPCA, respectively; \citet{sri}, DCLS-SPCA, \citet{she}, sPCA-rSVD, and \citet{jou}, \emph{Gpower}; \citet{fra}, BB-sPCA; \citet{wan} who proposed a simple but effective greedy backward elimination algorithm, SPCA-IE, for the same problem. Lastly, \citet{tre13} offers a number of solutions based on different approximations of the covariance matrix under $\Li$ constraints.

It should be noted that the comparisons are useful only to gain a feel of how different SPCA solutions compare to the LS-SPCA BB and BE algorithms. In fact, we showed that theoretically the LS-SPCA solutions are the ones that maximise the variance explained but the greedy algorithms may not find the global optima. It should also be considered that the BE algorithm is designed to attain large loadings at the cost of explaining less variance. Furthermore, our algorithms are implemented in R without optimising the code. Therefore scalability or speed issues will not be addressed. In spite of this we were able to compute sparse solutions for problems with up to almost 1000 variables in few minutes on a small computer.
\subsubsection{Performance measures}
Obviously, a performance measure of a set of SPCA solutions is the variance explained by each component, computed according to Equation (\ref{eq:vexp}), and the cumulated one. Hence we report the percentage of individual and cumulated variance explained, denoted as PVE and PCVE, respectively. The variance explained by the full PCs is an upper bound for the variance explained by the sparse components, hence we also consider the cumulated variance explained relative to that of PCA, denoted as PRCVE. A crucial feature of the sparse solutions is the cardinality of the loadings, we report these denoted as Card. Finally, we also report the absolute value of the smallest non-zero loading denoted as MinLoad, sometime expressed as a percentage contribution,  $\text{Min PCont} = 100\times|a_{ij}|/\sum_i |a_{ij}|$.
\subsubsection{Zou's synthetic data}\label{sec:zou}
\citet{zou} generated an artificial $10\times 10$ covariance matrix with underling dimension of two starting from three hidden variables:
$$ V_1 =  N( 0, 290),\: V_2 = N(0, 300),\: V_3 = -0.3 V_1 + 0.925 V_2 + \epsilon.$$
where $\epsilon$ is a Standard Normal variable. The manifest variables were then generated as
\begin{equation*}
X_i = V_1 + \epsilon_i,\; i = 1, 2, 3, 4;\;
X_i = V_2 + \epsilon_i,\; i = 5, 6, 7, 8;\;
X_i = V_3 + \epsilon_i,\; i = 9, 10
\end{equation*}
where the $\epsilon_i$'s are independent Standard Normal variables.
From the correlation matrix among the variables shown in Table \ref{tab:zoucov} it is easy to see that there exist three blocks of variables.
% Table generated by Excel2LaTeX from sheet 'Sheet1'
\begin{table}[H]
  \centering
  \caption{Correlations between variables in Zou's synthetic data.}
  {\scriptsize
\begin{tabular}{lrrrrrrrrrr}
\toprule
      & $x_1$ & $x_2$ & $x_3$ & $x_4$ & $x_5$ & $x_6$ & $x_7$ & $x_8$ & $x_9$ & $x_{10}$ \\
\midrule
$x_1$ & 1     & 0.996 & 0.996 & 0.996 & 0     & 0     & 0     & 0     & -0.3  & -0.3 \\
$x_2$ & 0.996 & 1     & 0.996 & 0.996 & 0     & 0     & 0     & 0     & -0.3  & -0.3 \\
$x_3$ & 0.996 & 0.996 & 1     & 0.996 & 0     & 0     & 0     & 0     & -0.3  & -0.3 \\
$x_4$ & 0.996 & 0.996 & 0.996 & 1     & 0     & 0     & 0     & 0     & -0.3  & -0.3 \\
$x_5$ & 0     & 0     & 0     & 0     & 1     & 0.997 & 0.997 & 0.997 & 0.95  & 0.95 \\
$x_6$ & 0     & 0     & 0     & 0     & 0.997 & 1     & 0.997 & 0.997 & 0.95  & 0.95 \\
$x_7$ & 0     & 0     & 0     & 0     & 0.997 & 0.997 & 1     & 0.997 & 0.95  & 0.95 \\
$x_8$ & 0     & 0     & 0     & 0     & 0.997 & 0.997 & 0.997 & 1     & 0.95  & 0.95 \\
$x_9$ & -0.3  & -0.3  & -0.3  & -0.3  & 0.95  & 0.95  & 0.95  & 0.95  & 1     & 0.948 \\
$x_{10}$ & -0.3  & -0.3  & -0.3  & -0.3  & 0.95  & 0.95  & 0.95  & 0.95  & 0.948 & 1 \\
\bottomrule
\end{tabular}%
 }
  \label{tab:zoucov}%
\end{table}%

Table \ref{tab:zou} shows the Varimax rotation of the first three PCA loadings together with the LS-SPCA(BB) solutions with the same cardinality, The last two columns are the LS-CSPCA correlated solutions with cardinality of one.
The SPCA methods \emph{SPCA}, DSPCA, SCoTLASS, BB-sPCA and SPCA-IE all find the first two sparse solutions that are equal to the varimax rotation of the PCA loadings. These solutions clearly identify the block structure of the data but they are not optimal neither with respect to the variance explained nor to sparsity. In fact, the LS-SPCA solutions of the same cardinality explain efficiently more variance. Furthermore, most of these methods do not require \uns\ of the components. The correlated LS-CSPCA solutions in the same table are much more parsimonious, efficient and interpretable solutions. Of course, on such small matrix the difference in variance explained is relatively small but on larger problems it would  be much larger.
\begin{table}[H]
 % Table generated by Excel2LaTeX from sheet 'Zou'
 \centering
  \caption{Varimax rotated PCA loadings, uncorrelated LS-SPCA(BB) loadings with same cardinality and correlated LS-CSPCA(BB) loadings with cardinality (1,1).}
   {\scriptsize % Table generated by Excel2LaTeX from sheet 'Zou Small'
\begin{tabular}{lrrrrrrrrr}
\toprule
      & \multicolumn{3}{c}{Varimax and SPCA} &       & \multicolumn{2}{c}{LS-SPCA--BB} &       & \multicolumn{2}{c}{CLS-SPCA--BB (1,1)}  \\
 \cmidrule{2-4}\cmidrule{6-7}\cmidrule{9-10}
      & Comp 1 & Comp 2 & Comp 3 &       & Comp 1 & Comp 2 &       & Comp 1 & Comp 2 \\
  \cmidrule{2-4}\cmidrule{6-7}\cmidrule{9-10}
$X_1$ &       & 0.5   &       &       &       & 0.516 &       &       &  \\
$X_2$ &       & 0.5   &       &       &       & 0.516 &       &       &  \\
$X_3$ &       & 0.5   &       &       &       & 0.516 &       &       &  \\
$X_4$ &       & 0.5   &       &       & 0.312 &       &       &       & 1 \\
$X_5$ & 0.5   &       &       &       &       &       &       &       &  \\
$X_6$ & 0.5   &       &       &       &       &       &       &       &  \\
$X_7$ & 0.5   &       &       &       & 0.536 & -0.45 &       &       &  \\
$X_8$ & 0.5   &       &       &       & 0.536 &       &       &       &  \\
$X_9$ &       &       & 0.71  &       &       &       &       &       &  \\
$X_{10}$ &       &       & 0.71  &       & 0.572 &       &       & 1     &  \\
\midrule
PVE  & 58.2  & 41.3  & 0.1   &       & 60.0  & 39.6  &       & 59.8  & 39.5 \\
PCVE & 58.2  & 99.4  & 99.8  &       & 60.0  & 99.6  &       & 59.8  & 99.3 \\
PRCVE & 96.9  & 99.7  & 100   &       & 99.9  & 99.9  &       & 99.5  & 99.6 \\
Card  & 4     & 4     & 2     &       & 4     & 4     &       & 1     & 1 \\
MinLoad & 0.5   & 0.5   & 0.71  &       & 0.312 & 0.451 &       & 1     & 1 \\
  \bottomrule
\end{tabular}%
   }
  \label{tab:zou}%
\end{table}

SPCA methods are designed to explain the most possible variance of the data but sometimes authors speak of block identification and model selection; for example, \citet{dasp08} build their method for explaining the most variance with \emph{statistical fidelity} but then venture into applying the sparse components to regression model selection. Then it is not clear why identifying data blocks or explaining an exogenous variable should be a feature of the solutions. In the literature can be found methods for sparse regression, for example.
\subsubsection{Pitprops data}\label{sec:pit}
\begin{figure}[H]
\centering
%C:/Users/giovanni/Documents/My Dropbox/Papers
\framebox{
    \includegraphics[width = 0.5\textwidth, keepaspectratio=True]{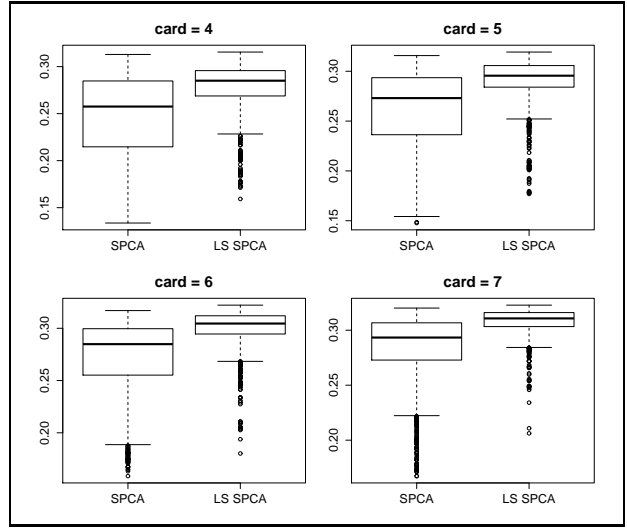}  %%width = 7 , height = 7cm
    }
\caption{distribution of the variance explained by the first PCs of the principal covariance submatrices and LS-SPCA solutions for all possible combinations of cardinality from four to seven.}
\label{fig:pitdist}
\end{figure}

The Pitprops dataset was used by \citet{jef} to illustrate the difficulty of interpreting PCs and has become a standard benchmark for SPCA methods. It consists of the correlation matrix of thirteen measures taken on pitprops selected according to a sampling design. Most of the existing SPCA methods were tested on this dataset. The small size of this problem does not allow for extensive comparisons and the differences in variance explained are not very large. This can be appreciated by observing the summary statistics  of the distribution of the variance explained by the PCs of first PCs of the principal covariance submatrices and LS-SPCA solutions for all possible combinations of cardinality from four to seven, shown in Figure \ref{fig:pitdist}. As expected, for every cardinality a large proportion of LS-SPCA solutions explain more variance than the PCs of the covariance submatrices. However, the maximum variance explained by the PCs is close to that of the LS-SPCA solutions. Hence, small differences are relevant for this example.
%Table \ref{tab:pitallvexp}.
Unfortunately, different methods were tested with different cardinalities, so it is impossible to compare each one with the others. Thus, we compare our methods with each one of them using the same cardinalities of the published results. Table \ref{tab:pitcomp} compares the results obtained with LS-CSPCA(BB), LS-SPCA(BB) and LS-CSPCA(BE) with those of other SPCA methods found in the literature, which are:
SPCA-IE, GSPCA and ESPCA, Gpower, BB-sPCA (two different cardinalities, "7" and "6"), \emph{SPCA}, DSPCA (two different cardinalities, "6" and "7"), SCoTLASS, DC-PCA and sPCA-rSVD. The results for GSPCA, Gpower and DC-PCA are grouped together under the label "GGD" because they are identical. Some of the LS-SPCA components could not be computed because the cardinality was too low to enforce \uns.
\begin{table}[H]
  \centering
  \caption{Pitprops data: comparison of the cumulative variance explained by different methods.}
  {\scriptsize
% Table generated by Excel2LaTeX from sheet 'AllTable PCVE'
\begin{tabular}{lrrrrrrrrrr}
\toprule
{\tiny Method}  &	 {\tiny ESPCA} &	 {\tiny SPCA-IE} &	 {\tiny GGD} &	 {\tiny DSPCA6} &	 {\tiny SCoTLASS} &	 {\tiny BB-sPCA6} &	 {\tiny DSPCA7} &	 {\tiny rSVD} &	 {\tiny SPCA} &	 {\tiny BB-sPCA7} \\
Card. &	 5 2 2 &	 6 2 2  &	 6 2 2  &	 6 2 3  &	 6 6 7 8 &	 6 7 7 8 &	 7 2 3  &	 7 2 4 7 &	 7 4 4 1 &	 7 4 4 1 \\
\midrule										
Comp. &	 &	&	 &	 &	&	&	&	 &	 &	 \\
1st &	 31.0 &	 31.3 &	 31.3 &	 30.7 &	 30.9 &	 31.3 &	 31.5 &	 31.8 &	 30.4 &	 31.9 \\
2nd &	 46.9 &	 47.5 &	 47.5 &	 46.4 &	 47.5 &	 48.8 &	 47.4 &	 47.8 &	 46.6 &	 48.9 \\
3rd &	 59.4 &	 61.0 &	 60.1 &	 59.7 &	 63.5 &	 63.4 &	 60.1 &	 62.8 &	 61.9 &	 57.9 \\
4th &	 &	&	 &	&	 71.0 &	 71.8 &	 &	 71.9 &	 70.2 &	 68.0 \\
\midrule										
 \multicolumn{10}{c}{LS-CSPCA(BB)}  \\										
1st &	 31.9 &	 32.2 &	 32.2 &	 32.2 &	 32.2 &	 32.2 &	 32.3 &	 32.3 &	 32.3 &	 32.3 \\
2nd &	 48.3 &	 48.7 &	 48.7 &	 48.7 &	 50.2 &	 50.3 &	 48.7 &	 48.7 &	 49.9 &	 49.9 \\
3rd &	 60.9 &	 61.3 &	 61.3 &	 62.3 &	 64.5 &	 64.7 &	 62.4 &	 63.0 &	 63.6 &	 63.6 \\
4th &	 &	&	 &	&	 73.2 &	 73.2 &	 &	 71.6 &	 71.6 &	 71.6 \\
 \midrule										
 \multicolumn{10}{c}{LS-SPCA(BB)} \\										
1st &	 31.9 &	 32.2 &	 32.2 &	 32.2 &	 32.2 &	 32.2 &	 32.3 &	 32.3 &	 32.3 &	 32.3 \\
2nd &	 48.2 &	 48.4 &	 48.4 &	 48.4 &	 50.2 &	 50.3 &	 48.5 &	 48.5 &	 49.8 &	 49.8 \\
3rd &	 &	&	 &	 60.7 &	 64.5 &	 64.7 &	 60.8 &	 62.1 &	 63.4 &	 63.4 \\
4th &	 &	&	 &	&	 73.2 &	 73.2 &	 &	 71.1 &	 &	\\
\midrule										
 \multicolumn{10}{c}{LS-CSPCA(BE)}  \\										
1st &	 31.6 &	 32.0 &	 32.0 &	 32.0 &	 32.0 &	 32.0 &	 32.3 &	 32.3 &	 32.3 &	 32.3 \\
2nd &	 47.9 &	 48.2 &	 48.2 &	 48.2 &	 49.9 &	 50.1 &	 48.7 &	 48.7 &	 49.8 &	 49.8 \\
3rd &	 60.5 &	 59.7 &	 59.7 &	 61.1 &	 64.2 &	 64.4 &	 62.3 &	 63.0 &	 63.5 &	 63.5 \\
4th &	 &	&	 &	&	 72.8 &	 73.0 &	 &	 71.6 &	 71.7 &	 71.7 \\
\bottomrule
\end{tabular}%
}
  \label{tab:pitcomp}%
\end{table}%
The LS-CSPCA(BB) solutions consistently explain more variance than any of the other methods The only case in which they explain less variance than an SPCA method is for the fourth dimension of rSVD. The uncorrelated LS-SPCA(BB) components explain less variance than the LS-CSPCA ones, but, still, more than those of the SPCA methods. The BE correlated algorithm perform worse than both BB solutions but also performs well when compared with the other methods, explaining less variance then rSVD, SPCA-IE and the GGD group only with the fourth component. Comparing the variance explained by the first compnents, it is evident that variance explained by the SPCA methods is never close to the optimum (the LS-SPCA solutions) or even to the suboptimum provided by the BE solutions.

Table \ref{tab:pitl2} compares the variance ($\Lt$ norms) of the first components computed with various SPCA methods with those of the corresponding LS-SPCA solutions. In every case the LS-SPCA components have smaller $\Lt$ norm but explain more variance, thus verifying that components with larger $\Lt$ norm do not necessarily explain the most variance.
%NEW Table
% Table generated by Excel2LaTeX from sheet 'Length comps'
\begin{table}[H]
  \centering
  \caption{Pitprops data: $\Lt$ norms of the first components computed with various SPCA methods and with LS-SPCA using the same cardinality.}
  {\scriptsize
        \begin{tabular}{lrrrrrrrrrr}
        \toprule
Method&  	 {\tiny ESPCA} &	 {\tiny SPCA-IE} &	 {\tiny GGD} &	 {\tiny DSPCA6} &	 {\tiny SCoTLASS} &	 {\tiny BB-sPCA6} &	 {\tiny DSPCA7} &	 {\tiny rSVD} &	 {\tiny SPCA}  &	{\tiny BB-sPCA7} \\
\midrule										
cardinality 	&5	& 6     	& 6 	& 6     	& 6     	& 6     	& 7     	 & 7     	 & 7  	 & 7\\
\midrule										
All methods 	&3.41	& 3.74  	& 3.74 	& 3.46  	& 3.71  	& 3.77  	& 3.82  	 & 3.99  	 & 3.64   	 & 4.00\\
LS-SPCA 	&2.29 	& 2.78  	& 2.78	& 2.78  	& 2.78  	& 2.78  	& 3.28  	 & 3.28  	 & 3.28  	 &3.28\\
\bottomrule
\end{tabular}%
}
  \label{tab:pitl2}%
\end{table}%
\citet{tre13} presents results for the Pitprops data for various new SPCA solutions with yet different cardinalities. These methods optimize different approximations of the correlation matrix under $\Li$ constraints, discussing which is beyond the scope of this paper. Therefore, we will refer to each method as T$n\, \mu=m$, where $n$ refers to the reference number of the formula defining the objective function in the original paper and $\mu$ is the value of the tuning parameter used. The T10 method computes \un\ components, therefore we used LS-SPCA(BB) for this comparison and LS-CSPCA(BB) for all other cases.

Table \ref{tab:pittre} compares the summary results of \citeauthor{tre13} with the corresponding LS-SPCA(BB) ones.
Clearly, also in this case the LS-SPCA components explain more variance, most markedly with the first few ones. The only exception is the second component of the T8 method, which explains slightly more variance than the corresponding LS-CSPCA one. However, the following components perform decidedly worse. Note that, in most cases, \citeauthor{tre13}'s solutions present small loadings in the first components.
% Table generated by Excel2LaTeX from sheet 'Sheet6'
\begin{table}[H]
  \centering
  \caption{Pitprops data: \citeauthor{tre13}'s solutions compared with the corresponding LS-CSPCA(BB) ones.}
  {\scriptsize
\begin{tabular}{lrrrrrrrrrrrrr}
\toprule
Component & C1 & C2 & C3 & C4 & C5 & C6 & & C1 & C2 & C3 & C4 & C5 & C6 \\
\midrule
Method & \multicolumn{6}{c}{T5 $\mu=5$, Card  (10  6  3  1  3  2) } & & \multicolumn{6}{c}{LS-CSPCA(BB)}\\
\cmidrule{2-7}\cmidrule{9-14}
%PCVE & 32.2 & 47.2 & 62.4 & 70.6 & 76.9 & 83.6 & & 32.4 & 50.4 & 63.6 & 71.6 & 78.5 & 85.1 \\
PRCVE & 99.3 & 93 & 95.7 & 95.7 & 95.3 & 96.1 & & 100 & 99.3 & 97.6 & 97.1 & 97.3 & 97.9 \\
MinLoad & 0.017 & 0.2 & 0.322 & 1 & 0.024 & 0.272 & & 0.110 & 0.147 & 0.458 & 1 & 0.415 & 0.550 \\
\midrule
Method & \multicolumn{6}{c}{T7 $\mu=6$,Card  (7  2  3  1  1  1) } & & \multicolumn{6}{c}{LS-CSPCA(BB)}\\
\cmidrule{2-7}\cmidrule{9-14}
%PCVE & 31.1 & 46.6 & 59.4 & 66.4 & 74.4 & 80.8 & & 32.3 & 48.7 & 62.4 & 70.7 & 77.6 & 84.1 \\
PRCVE & 95.7 & 91.9 & 91.1 & 90.1 & 92.2 & 92.9 & & 99.5 & 96.1 & 95.7 & 96 & 96.1 & 96.7 \\
MinLoad & 0.059 & 0.707 & 0.122 & 1 & 1 & 1 & & 0.289 & 0.467 & 0.418 & 1 & 1 & 1 \\
\midrule
Method & \multicolumn{6}{c}{T8 $\mu=3$, Card  (7  1  3  2  1  1)} & & \multicolumn{6}{c}{LS-CSPCA(BB)}\\
\cmidrule{2-7}\cmidrule{9-14}
%PCVE & 30 & 41.6 & 50.5 & 64.6 & 72.5 & 79.2 & & 32.3 & 41.4 & 58 & 70.4 & 76.1 & 83.1 \\
PRCVE & 92.5 & 82 & 77.5 & 87.6 & 89.8 & 91 & & 99.5 & 81.5 & 89 & 95.5 & 94.3 & 95.5 \\
MinLoad & 0.081 & 1 & 0.432 & 0.707 & 1 & 1 & & 0.289 & 1 & 0.418 & 0 & 1 & 1 \\
\midrule
Method & \multicolumn{6}{c}{T9 $\mu=4$, Card   (7      2      1      1      1      1)} & & \multicolumn{6}{c}{LS-CSPCA(BB)}\\
\cmidrule{2-7}\cmidrule{9-14}
%PCVE & 31.6  & 46.9  & 55.2  & 62.5  & 72.1  & 80.4  &       & 32.3  & 48.7  & 57.2  & 67.4  & 76.1  & 82.6 \\
PRCVE & 97.3  & 92.3  & 84.7  & 84.7  & 89.4  & 92.4  &       & 99.5  & 96.1  & 87.7  & 91.4  & 94.2  & 94.9 \\
MinLoad & 0.102 & 0.708 & 1     & 1     & 1 & 1 &       & 0.289 & 0.467 & 1 & 1     & 1 & 1 \\
\midrule
Method & \multicolumn{6}{c}{T10 $\mu=5$, Card  (7  6  4  4  5  6)} & & \multicolumn{6}{c}{LS-SPCA(BB)}\\
\cmidrule{2-7}\cmidrule{9-14}
%PCVE & 31.7 & 47 & 63 & 71.3 & 78.6 & 84.9 & & 32.3 & 50.2 & 63.9 & 72.2 & 79.1 & 85.5 \\
PRCVE & 97.6 & 92.6 & 96.6 & 96.7 & 97.4 & 97.5 & & 99.5 & 99 & 97.9 & 98 & 98 & 98.3 \\
MinLoad & 0.132 & 0.186 & 0.146 & 0.059 & 0.031 & 0.063 & & 0.289 & 0.152 & 0.320 & 0.14296 & 0.137 & 0.048 \\
\midrule
Method & \multicolumn{6}{c}{T11 $\mu=4$, Card  (7  4  1  1  2  3)} & & \multicolumn{6}{c}{LS-CSPCA(BB)}\\
\cmidrule{2-7}\cmidrule{9-14}
%PCVE & 31.3 & 47.7 & 56 & 66.5 & 74.9 & 81.4 & & 32.3 & 49.9 & 59.8 & 67.9 & 77.6 & 84.4 \\
PRCVE & 96.3 & 93.9 & 86 & 90.2 & 92.8 & 93.6 & & 99.5 & 98.4 & 91.7 & 92 & 96.1 & 97 \\
MinLoad & 0.016 & 0.059 & 1 & 1 & 0.087 & 0.091 & & 0.289 & 0.234 & 1 & 1 & 0.530 & 0.434 \\
\bottomrule
\end{tabular}%
}
 \label{tab:pittre}%
\end{table}%
\subsection{Examples}
In this section we illustrate some results obtained by applying LS-SPCA to datasets of different dimensionality available in the literature or on the StatLib Data Archive (\emph{lib.stat.cmu.edu}) and UCI Machine Learning
Repository (\emph{archive.ics.uci.edu}, \citealp{uci}). Since there is a trade-off between explaining the variance and sparsity, the results are not necessarily the "best" ones with respect to either requirements. We present what we obtained with what we consider reasonable working requirements for illustrative purposes.
\subsubsection*{Anthropometric measures}
The oldest application of PCA is the analysis of a set of seven anthropometric measures taken on a sample of non-habitual criminals \citep{macd}. The data were used to classify criminals by their physical features. We use this small dataset to illustrate the differences among the LS-SPCA solutions obtained with the BB and BE algorithms and the globally optimal ones. We obtained the global optima by exploring all possible solutions for three components with cardinality (2, 2, 3) and (2, 3, 3). The summary results are shown in Table \ref{tab:anth}. As expected, the first BB components explain more total variance of all methods but the complete search attains a higher global optimum. The slightly higher PCVE of the first solution results in a loss of PCVE when other components are included; for cardinalities (2, 2, 3) the two BB components still perform better than the global optimum, but not for cardinalities (2, 3, 3).
The BE algorithm performs notably worse than the BB one for the cardinality (2, 2, 3) but in the other case it outperforms it by finding the optimal solutions. In this case the BB search pays for its initial greediness by getting stuck in a local maximum while BE does not.

This is the only example we observed in which BE outperforms BB in early components, other cases were observed for a larger number of components and by small differences. In our observations, the BE solutions are usually not much worse than the BB ones.
% Table generated by Excel2LaTeX from sheet 'Anthrop'
\begin{table}[H]
  \centering
  \caption{Anthropometric measures: summary results of the globally optimal, BB and BE solutions with cardinalities (2, 2, 3) and (2, 3, 3).}
  {\scriptsize
\begin{tabular}{lrrrrrrrrrrr}
\toprule
Method & \multicolumn{3}{c}{Optimal} &       & \multicolumn{3}{c}{BB} &       & \multicolumn{3}{c}{BE}  \\
\midrule
      & Comp 1 & Comp 2 & Comp 3 &       & Comp 1 & Comp 2 & Comp 3 &       & Comp 1 & Comp 2 & Comp 3 \\
      \cmidrule{2-4}\cmidrule{6-8}\cmidrule{10-12}
      & \multicolumn{11}{c}{Three components, cardinality 2 2 3} \\
\midrule
PVE   & 49.2  & 13.6  & 18.7  &       & 49.3  & 18.7  & 10.9  &       & 49.2  & 11.6  & 12.6 \\
PCVE  & 49.2  & 62.7  & \textbf{81.5} &       & \textbf{49.3} & \textbf{67.9} & 78.8  &       & 49.2  & 60.8  & 73.4 \\
PRCVE & 90.6  & 82.9  & 95.8  &       & 90.8  & 89.7  & 92.7  &       & 90.6  & 80.3  & 86.3 \\
Min PCont & 44.4  & 34.2  & 15.4  & 0     & 29.7  & 26.1  & 17.3  &       & 44.4  & 34.1  & 10.1 \\
      & \multicolumn{11}{c}{Three components, cardinality 2 3 3}                               \\
\midrule
PVE   & 49.2  & 23.2  & 10.7  &       & 49.3  & 20.4  & 11.2  &       & 49.2  & 23.2  & 10.7 \\
PCVE  & 49.2  & \textbf{72.3} & \textbf{83} &       & \textbf{49.3} & 69.7  & 80.9  &       & 49.2  & \textbf{72.3} & \textbf{83} \\
PRCVE & 90.6  & 95.5  & 97.6  &       & 90.8  & 92    & 95.1  &       & 90.6  & 95.5  & 97.6 \\
Min PCont & 44.4  & 20.8  & 12.8  & 0     & 29.7  & 27.5  & 8.2   &       & 44.4  & 20.8  & 12.8 \\
\bottomrule
\end{tabular}%
}
  \label{tab:anth}%
\end{table}%
The results in this example can be replicated using the Anthrop dataset included in the R package \emph{spca}.
\subsubsection*{Baseball hitters}
This dataset (available at Statlib) contains observations on 16 performance statistics of 263 US Major League baseball hitters taken on their career and on 1986 season. The data were used to explain the players'salary.

We computed five components with the BE algorithm, trimming the percent contributions to the thresholds 0.35, 0.35, 0.2, 0.2 and 0.2, respectively. All the components but the fourth one could be trimmed to the required threshold; for the fourth one trimming ended because the minimal cardinality to permit \uns\ was reached. The summary results are shown in Table \ref{tab:hit1} together with those of the BB solutions computed with the same cardinalities. The latter method explains slightly more variance but the last two components present a percent contribution of three percent or less.
% Table generated by Excel2LaTeX from sheet 'bsbl'
\begin{table}[H]
  \centering
  \caption{Baseball Hitters data: comparison of LS-SPCA(BB) and LS-SPCA(BE) results.}
  {\scriptsize
  \begin{tabular}{lrrrrrrrrrrr}
\toprule
      & \multicolumn{5}{c}{BE} & & \multicolumn{5}{c}{BB}\\
\cmidrule{2-6}\cmidrule{8-12}
Statistics& Comp1 & Comp2 & Comp3 & Comp4 & Comp5 &       & Comp1 & Comp2 & Comp3 & Comp4 & Comp5 \\
\midrule
PVE   & 44.5  & 24.6  & 10.9  & 5.7   & 4.1   &       & 44.5  & 24.7  & 10.8  & 5.7   & 4.4 \\
PCVE  & 44.5  & 69.1  & 79.9  & 85.6  & 89.7  &       & 44.5  & 69.2  & 80.1  & 85.7  & 90.1 \\
PRCVE & 98.2  & 97.3  & 97.7  & 98.1  & 98    &       & 98.2  & 97.5  & 97.9  & 98.3  & 98.3 \\
Card  & 3     & 3     & 4     & 4     & 7     &       & 3     & 3     & 4     & 4     & 7 \\
MinPContr & 24.8  & 27.5  & 13.4  & 8.1   & 11.1  &       & 24.5  & 18.0    & 14.4  & 3.0     & 1.7 \\
\bottomrule
\end{tabular}%
}
  \label{tab:hit1}%
\end{table}%
The results in this example can be replicated using the Anthrop dataset included in the R package \emph{spca}.
\subsubsection*{Optical Recognition of Handwritten Digits}
The optdigits dataset (available at the UCI Repository) contains measures on graphical attributes of different handwritten digits, which were used to classify the digits. We merged the training and test samples and removed the classification variable and two constant ones, which left 62 variables. We ran the BE algorithm requiring that the variance explained by each component was at least 90\% of that explained by the initial untrimmed solution (with reference to Algorithm \ref{algo:spcabe}, we set the \emph{mvl} threshold to 0.1). Table \ref{tab:opt} shows the summary results for the first five components. Each solution gives over 91\% of PRCVE with at most 11 loadings.
%odc.spbe1v = spcabe(odc, nd = F, thresh = F, threshvar = 0.1, ndbyvexp = 0.6)
% Table generated by Excel2LaTeX from sheet 'Optrecog'
\begin{table}[H]
  \centering
  \caption{Optical Recognition data: LS-SPCA(BE) summary results.}
  {\scriptsize
\begin{tabular}{lrrrrr}
\toprule
      & Comp1 & Comp2 & Comp3 & Comp4 & Comp5\\
\midrule
PVE   & 10.7  & 9.2   & 7.1   & 5.4   & 4.5 \\
PCVE  & 10.7  & 19.9  & 27.0  & 32.4  & 36.9 \\
PRCVE & 92.1  & 91.3  & 91.3  & 91.5  & 91.7 \\
Card  & 6     & 7     & 7     & 10    & 11 \\
Min PContr & 14.2  & 9.8   & 10.3  & 7.4   & 6.0 \\
\bottomrule
\end{tabular}%
}
  \label{tab:opt}%
\end{table}%
\subsubsection*{US crime data}
This dataset (available at the UCI repositiory) contains socioeconomic records on different US cities collected in the 90s. The data was used to explain the rate of violent crime in each city. We deleted 22 variables and one observation with missing values. The final set contains 1994 observations on 99 variables. We run LS-SPCA(BE) requiring that the cumulative variance explained after including each component was at least 95\% of that explained by the PCs. The summary results of first five components are shown in Table \ref{tab:crime}.
The solutions explain over 95\% of the variance explained by the ordinary PCs with extremely low cardinality and all contributions above 11\%.
% Table generated by Excel2LaTeX from sheet 'CrimeUS'
\begin{table}[H]
  \centering
  \caption{US crime data: LS-SPCA(BE) summary results.}
  {\scriptsize
  \begin{tabular}{lrrrrr}
\toprule
      & Comp1 & Comp2 & Comp3 & Comp4 & Comp5 \\
\midrule
% Table generated by Excel2LaTeX from sheet 'CrimeUS'
PVE   & 24.1  & 16.4  & 8.7   & 7.3   & 5.6 \\
PCVE  & 24.1  & 40.5  & 49.2  & 56.5  & 62.1 \\
PRCVE & 95.4  & 95.8  & 95.3  & 95.4  & 95.6 \\
Card  & 2     & 4     & 3     & 6     & 6 \\
MinPContr & 40.6  & 13.1  & 25.3  & 11.7  & 11.3 \\
\bottomrule
\end{tabular}%
    }  \label{tab:crime}%
\end{table}%
\subsubsection*{One hundred plant species leaves dataset}
The \emph{100 leaves plant species} dataset (available at the UCI repositiory) was obtained by merging the measurements on three different aspects of the shapes of leaves of one hundred different species of plants. It was used for classifying the plants.
We removed one observation from two files because it was missing from the other. So, the final set contained 1599 instances of 186 variables. We ran LS-SPCA(BE) requiring that each of the first five component computed explained at least 90\% of the variance explained by the initial untrimmed solution.
The summary results are shown in Table \ref{tab:leaves}. The PRCVE for first component is 98\% with just two loadings, the addition of the second component leads to a PRCVE of almost 96\%, with a total of just seven loadings, and for the first three PRCVE is 95.9\%, with just a total of 11 loadings. The following components have higher cardinality and also explain over 95\% of the variance explained by the PCs.
%le.spc4 = spcabe(le.cor, nd = F, thresh = F, threshvar = 0.1, ndbyvexp = 0.7)
\begin{table}[H]
  \centering
  \caption{Plant Leaves data: LS-SPCA(BE) summary results.}
  {\scriptsize
% Table generated by Excel2LaTeX from sheet 'LeavesAll'
\begin{tabular}{lrrrrr}
\toprule
      & Comp1 & Comp2 & Comp3 & Comp4 & Comp5\\
\midrule
PPVE   & 23.7\% & 9.5\% & 6.0\% & 5.0\% & 3.8\% \\
PCVE  & 23.7\% & 33.3\% & 39.3\% & 44.3\% & 48.1\% \\
PRCVE & 97.9\% & 95.9\% & 95.4\% & 95.3\% & 95.1\% \\
Card  & 2     & 5     & 4     & 10    & 12 \\
Min PContr & 49.1  & 7.2   & 16.6  & 7.1   & 5.5 \\
\bottomrule
\end{tabular}%
    }
  \label{tab:leaves}%
\end{table}%
\subsubsection*{Isolated Letter Speech Recognition}
The \emph{isolet1+2+3+4} dataset (available at the UCI Repository) contains 617 measurements taken on the sounds produced by readers speaking the name of different letters of the alphabet. The first eight PCs explain 50.4\% of the total variance. Therefore, we ran LS-SPCA(BE) requiring not more than 10 components or up to 50\% of total variance explained. For the first three components trimming was stopped if the loss of PVE from the initial component was greater than 10\%, for the next three components this value was 15\% and 20\% for the remaining ones. The algorithm took 158 minutes to terminate with 10 components and 48.8\% total variance explained. The summary results are shown in Table \ref{tab:isolet}. The first component explains 93.7\% of the variance explained by the first PC with just three non-zero loadings. The addition of the following components leads to explaining decreasing proportions of the variance explained by the PCs, but always around 90\% of it. The loadings have at most cardinality of ten and only one is smaller than 0.1.
\begin{table}[H]
  \centering
  \caption{Speech recognition data: LS-SPCA(BE) results.}
  {\scriptsize
% Table generated by Excel2LaTeX from sheet 'Sheet4'
\begin{tabular}{lrrrrrrrrrr}
\toprule
      & Comp1 & Comp2 & Comp3 & Comp4 & Comp5 & Comp6 & Comp7 & Comp8 & Comp9 & Comp10 \\
\midrule
PVE  & 18.1 & 8.2 & 5.0 & 3.9 & 3.6 & 2.5 & 2.3 & 1.9 & 1.8 & 1.6 \\
PCVE & 18.1 & 26.2 & 31.2 & 35.1 & 38.7 & 41.2 & 43.5 & 45.4 & 47.2 & 48.8 \\
PRCVE & 93.7 & 93.1 & 92.8 & 92.1 & 91.9 & 90.9 & 90.6 & 90.0 & 89.8 & 89.5 \\
Card  & 3     & 10    & 6     & 9     & 10    & 8     & 9     & 9     & 9     & 10 \\
MinLoad & 0.438 & 0.233 & 0.303 & 0.256 & 0.227 & 0.092 & 0.210 & 0.126 & 0.192 & 0.238 \\
\bottomrule
\end{tabular}%
    }
  \label{tab:isolet}%
\end{table}%
% Table generated by Excel2LaTeX from sheet 'Sheet5'
\subsubsection*{Summary of the results}
Table \ref{tab:sum} shows the cumulative variance explained by the LS-SPCA(BE) components and their cardinalities. For all datasets there is a very consistent reduction of the cardinality while a large proportion of the variance explained by the PCs is preserved. As mentioned above, there is a trade-off between low cardinality and high proportion of variance explained, therefore these results are only indicative of a generic performance of the LS-SPCA(BE) algorithm and can be improved in favor of either feature.
\begin{table}[H]
  \centering
  \caption{PVE and cardinality of the LS-SPCA(BE) results for the different datasets considered.}
  {\scriptsize
% Table generated by Excel2LaTeX from sheet 'Summary table'
\begin{tabular}{lrrrrrrrrrr}
\toprule
      & Comp1 & Comp2 & Comp3 & Comp4 & Comp5 & Comp6 & Comp7 & Comp8 & Comp9 & Comp10 \\
\cmidrule{1-6}
Dataset and dimension & \multicolumn{5}{l}{Baseball Hitters, p = 13} &       &       &       &        \\
 PRCVE & 98.2  & 97.5  & 97.9  & 98.3  & 98.2  &       &       &       &       &  \\
 Cardinality & 3 & 3 & 4 & 4 & 5 &       &       &       &       &  \\
\midrule
Dataset and dimension & \multicolumn{10}{l}{Optical Recognition, p = 62}                              \\
 PRCVE & 92.1  & 91.3  & 91.3  & 91.5  & 91.4  & 92.1  & 92.1  & 92.2  & 92.7  & 93.1 \\
 Cardinality & 6 & 7 & 7 & 10 & 10 & 13 & 12 & 12 & 13 & 15 \\
\cmidrule{1-9}
Dataset and dimension & \multicolumn{8}{l}{US Crime, p = 99}                          &       \\
 PRCVE & 95.4 & 95.8 & 95.3 & 95.4 & 95.6 & 95.5 & 95.6 & 95.3 &       &  \\
 Cardinality & 2 & 4 & 3 & 6 & 6 & 8 & 8 & 9 &       &  \\
\midrule
Dataset and dimension & \multicolumn{10}{l}{100 Leaves. p = 186}                                      \\
 PRCVE & 97.9 & 95.9 & 95.4 & 95.3 & 95.1 & 94.6 & 94.3 & 94.4 & 94.4 & 94.2 \\
 Cardinality & 2 & 5 & 4 & 10 & 12 & 6 & 14 & 20 & 13& 15\\
\midrule
Dataset and dimension & \multicolumn{10}{l}{Isolated Letters, p = 617}                                \\
 PRCVE & 93.7 & 93.1 & 92.8 & 92.1 & 91.9 & 90.9 & 90.6 & 90.0 & 89.8 & 89.5 \\
 Cardinality & 3 & 10 & 6 & 9 & 10 & 8 & 9 & 9 & 9 & 10 \\
\bottomrule
\end{tabular}%
    }
  \label{tab:sum}%
\end{table}%
\subsubsection{Computational times}
We present a study on the computational time taken by LS-SPCA(BB) and LS-SPCA(BE)  algorithms. The elapsed time is
computed on a 64bit  quadri-core Intel i5 \textsuperscript{\textregistered} CPU with
and 4Gb RAM, using non-optimized R \citep{R} code.

Branch-and-bound searches are known to take exponential time; \citet{fra} shows how the BB algorithm becomes very time consuming as the dimension of the matrices increases. A comparison of the computing times taken by the LS-SPCA(BB) and LS-SPCA(BE) to compute solutions of increasing complexity on the Pitprops data is shown in Table \ref{tab:pittime}. Clearly the BE algorithm is much faster with a peak of 141 times shorter computational time. These results should be evaluated considering that the complexity for the BE algorithm decreases as the cardinality increases while for the BB one it is maximal when the cardinality is near half the dimension of the problem.
\begin{table}[H]
  \centering
  \caption{Pitprops data: comparison of the computational times taken by the BE and BB algorithms for different cardinalities. Time in seconds.}
  {\scriptsize
% Table generated by Excel2LaTeX from sheet 'Sheet1'
\begin{tabular}{lrrrr}
\toprule
%      &       & \multicolumn{3}{c}{Time} \\
%\cmidrule{3-5}
Cardinality & Replications & BE    & BB    & Relative \\
\midrule
2 2 3 4 5 & 30    & 4.43  & 128.17 & 28.932 \\
2 2 6 6 6 & 30    & 2.97  & 147.5 & 49.663 \\
6 6 3 4 5 & 30    & 4.49  & 259.44 & 57.782 \\
2 6 6 6 6 & 30    & 3.78  & 289.42 & 76.566 \\
6 6 6 6 6 & 30    & 1.95  & 275.41 & 141.236 \\
\bottomrule
\end{tabular}%
}
  \label{tab:pittime}%
\end{table}%
Table \ref{tab:compall} shows a comparison of the computational times taken by the BB and BE algorithms to compute 5 components of cardinality 10 on datasets of increasing dimension. The computational times increase exponentially with the dimension of the dataset.
\begin{table}[H]
  \centering
  \caption{Different datasets: comparison  of LS-SPCA(BE) computational times for different number of dimensions. Time in seconds.}
  {\scriptsize
% Table generated by Excel2LaTeX from sheet 'TimingFinal'
\begin{tabular}{lrrrrr}
\toprule
Dataset  & Dimension & Replications & Average & Relative & Sec per dimension \\
\midrule
Optical Digits & 62    & 100   & 1.1   & 1     & 0.02 \\
Crime in US & 99    & 100   & 4.7   & 4.2   & 0.05 \\
100 Leaves & 186   & 100   & 40.9  & 36.8  & 0.22 \\
Isolated Letters & 617   & 20    & 6766.4 & 6079.4 & 10.97 \\
\bottomrule
\end{tabular}%
}
  \label{tab:compall}%
\end{table}%
Finally, we compared the the computational time for trimming a different number of loadings at each iteration. The results of computing 5 components of cardinality 10 on the Isolated Letters dataset trimming 1, 5, 10, 20 and 50 loadings at the time are shown in Table \ref{tab:timetrim}. As expected, the computational time decreases as the number of loading trimmed increases; the relationship, for this example, is of log-log type.
%, as shown in Figure \ref{fig:trim}.
% Table generated by Excel2LaTeX from sheet 'TimingFinal'
\begin{table}[H]
  \centering
  \caption{Time taken to compute 5 components on the isolet dataset for increasing number of loadings trimmed at each iteration. Time in minutes.}
  {\scriptsize
% Table generated by Excel2LaTeX from sheet 'TimingFinal'
\begin{tabular}{rrrr}
\toprule
Trimmed & Replications & Average Time& Relative \\
\midrule
1     & 20    & 112.773 & 36.4 \\
5     & 20    & 24.8  & 8.0 \\
10    & 20    & 11.4  & 3.7 \\
20    & 20    & 5.6   & 1.8 \\
50    & 20    & 3.1   & 1 \\
\bottomrule
\end{tabular}%
}
  \label{tab:timetrim}%
\end{table}%
\section{Discussion}\label{sec:concl}
The popularity of PCA is due to its ability of summarising a set of variables with few components. SPCA greatly enhances the interpretability of these components.  SPCA-LS represents an improvement over other SPCA methods because it maintaining the PCs' key properties: \uns\ and LS optimality. The model based approach adopted overcomes the difficulties in defining deflations and variance explained existing for other methods.

The problem of simplifying the components has been widely discussed in the Factor Analysis literature. From the discussion it comes out clearly that there are different definitions and requirements for "simplicity", (for example, the ones suggested by \citealp{thu}), which cannot all be included in an objective function. In fact solving the PCA problem with restrictions on the cardinality achieves sparsity but does not guarantee large loadings or efficiency in explaining the variance. In this sense, the BE algorithm can be a useful tool for including more simplicity criteria into the problem by constraining the way the solutions are found.  In our implementation of the BE algorithm we include the possibility of excluding from a solution variables that are already in previous ones (so achieving parsimony) and the possibility of selecting in advance we variables can enter a component. Hence, the BE algorithm is very useful for selecting a "good" solution, that is one that is easy to interpret but that satisfies other requirements. This is very difficult to achieve with black-box numerical solvers, for as sophisticated and scalable as they could. Instead, using adaptable searches requirement like identifying blocks of data could be easily met. However, it could be useful to develop a more efficient computational algorithm for finding the LS-SPCA solutions on larger problems, of the kind existing for the maximisation of the variance of the components.\bibliographystyle{apalike}
\bibliography{SPCA}
\section*{Appendix}
\subsection*{Proof of the result}
The extra sum of squares principle says that the variance explained by an additional variable in a regression model is equal to variance explained by its orthogonal complement to the other variables. Consequently, the variance explained by a component computed under \uns\ constraints is a sup for the \vexp\ of correlated components.
\begin{proposition}[extra sum of squares principle]
A component correlated with the preceding ones cannot explain more variance than its complement orthogonal to the others.
\end{proposition}
\begin{proof}
Let $\bT =\bX\bA$ be a set of components and $\bz = \bX\bb$ be another component such that $\bz\trasp\bt_j \neq 0$ for at least one $j$. Also let $\bP = \bT(\bT\trasp\bT)^{-1}\bT\trasp$ be the orthogonal projector onto the space of $\bT$ and $\bQ = \bI - \bP$ its complement.

The variance explained by the components $\bT$ is $\Tr\left(\bX\trasp\bP\bX\right)$. We can write the components $\bG= [\bT;\bz]$ as
$\bG = \bP\bG + \bQ\bG = \bT\bM + \tilde \bz$, where $\bM$ is a matrix and $\tilde \bz = \bQ\bz$ is the complement of $\bz$ orthogonal to the $\bT$ components. Then, the regression of $\bX$ onto $\bG$ is $\hat \bX = \bG\left(\bG\trasp\bG\right)^{-1}\bG\bX = \bP\bX + \tilde\bz(\tilde \bz\trasp\tilde \bz)^{-1}\tilde \bz\trasp\bX$.
It follows that the variance explained by the components in $\bG$ is given by
$\Tr(\hat \bX\trasp\hat \bX) = \Tr(\bX\trasp\bP\bX) + \Tr(\bX\trasp\tilde\bz(\tilde \bz\trasp\tilde \bz)^{-1}\tilde \bz\trasp\bX)$. Therefore, the variance explained by the correlated component $\bz$ is equal to that explained by its orthogonal complement to the $\bT$ variables, completing the proof.
\end{proof}
\subsubsection*{Uncorrelated LS-SPCA components}
Here we derive the solutions for Problem (\ref{eq:uspcals}) based on a generalization of the proof for constrained multiple regression as given in \citep[Theorem 13.5]{mag}. We adopt the notation defined in Section \ref{sec:mod}, assuming that $\bD_j$ is invertible and that $\bR\trasp_j\bR_j$ is singular.
\begin{proposition}
The uncorrelated sparse principal components of given indices $ind_j$ that minimize the $\Lt$ norm of the residuals of the approximation are defined as
\begin{align}\label{eqa:uspcap}
        &\ba_j = \argmin\limits_{\ba_j \in \Re^{p}}
        ||\bX - \bX\ba_j\bp\trasp_j|| =
\argmax\limits_{\ba_j \in \Re^{p}}
\frac{\ba\trasp_j\bS\bS\ba_j}{\ba\trasp_j\bS\ba_j},\; j = 1,\ldots,d\\
&{\rm subject\ to}\;
        a_{i,j} = 0\ {\rm if}\ i \not\in ind_j\;
        {\rm and}\; \bR_j\ba_k=\mathbf{0},\; j < k.\nonumber
\end{align}
The solutions are given by the eigenvectors satisfying
\begin{equation}\label{proof:optspca}
\left[\bI_{n_j} - \bD_j^{-1} \bR\trasp_j(\bR_j\bD_j^{-1}\bR\trasp _j)^+\bR_j\right]\bB_j^*\bta_j = \bC_j\bB_j^*\bta_j = \phi_{\text{max}}\bta_j,
\end{equation}
where the subscript "+" denotes a generalized inverse, $\bC_j = [\bI_{n_j} -  \bD_j^{-1} \bR\trasp_j(\bR_j\bD_j^{-1}\bR\trasp_j)^+\bR_j]$, $\bC_1=\bI_{n_1}$, and $\bB_j^* = \bD_j^{-1}\bJ\trasp_j\bS\bS\bJ_j$.
The solutions exist because $\bR\trasp_j$ spans the space of $\bW\trasp_j$.
\end{proposition}
\begin{proof}
The objective function can be easily derived from the general problem in Equation (\ref{eq:lsspca}) by first observing that the $\bP\trasp$ matrix is a matrix of regression coefficients for the components $\bT = \bX\bA$. By developing the $\Lt$ norm as a trace and using the \uns\ of the components it is easy to obtain
\begin{align*}
        &\bA = \argmin\limits_{\ba_j \in \Re^{p\times d}}
        ||\bX - \bX\bA\bP\trasp|| =
\argmax\limits_{\ba_j \in \Re^{p}}
\sum_{j=1}^d\frac{\ba\trasp_j\bS\bS\ba_j}{\ba\trasp_j\bS\ba_j},\\
&{\rm subject\ to}\;
        a_{i,j} = 0\ {\rm if}\ i \not\in ind_j\;
        {\rm and}\; \bt_j\trasp\bt_k={0},\; j \neq k.\nonumber
\end{align*}
Because of the \uns\ the problem can be solved separately for each component as in Equation (\ref{eqa:uspcap}).

%Let $\bt_j = \bW_j\bta_j = \bX\bJ_j\bta_j$ be the $j$-th component.
By Problem (\ref{eqa:uspcap}), we want to maximize the variance explained subject to $\bR_j\bJ_j\bta_j = 0$. Hence, we need to maximize the Langrangian:
\begin{align}
L(\bta_j, \bla) &= \text{tr}\left[\bX\trasp\bt_j(\bt\trasp_j\bt_j)^{-1}\bt\trasp_j\bX\right] - 2\bla\trasp\bR_j\bta_j\nonumber\\
&= (\bta\trasp_j\bD_j\bta_j)^{-1}\bta\trasp_j\bJ\trasp_j\bS
\bS\bJ_j\bta_j - 2\bla\trasp\bR_j\bta_j.
\end{align}
Equating the partial derivatives to zero gives:
\begin{align}
\frac{\partial L}{\partial \bta} &= -\bD_j\bta_j \alpha_j +  \bJ\trasp_j\bS
\bS\bJ_j\bta_j\beta_j - \bR\trasp_j\bla = \mathbf{0}\label{eqa:part1}\\
\frac{\partial L}{\partial \bla} &= \bR_j\bta_j = \mathbf{0},\label{eqa:part2}
\end{align}
where
$\alpha_j = \bta\trasp_j\bJ\trasp_j\bS_j \bS_j\bJ_j\bta_j(\bta\trasp_j\bD_j\bta_j)^{-2}$ and $\beta_j = (\bta\trasp_j\bD_j\bta_j)^{-1}$. Since for the first component $\bR_1 = \mathbf{0}$, the solution is
\begin{equation}\label{eqa:proof01}
  \bD_1^{-1}\bJ\trasp_1\bS
\bS\bJ_1\bta_1  = \bta_1 \frac{\alpha_j}{\beta_j}.
\end{equation}
For the subsequent components we have that
premultiplying equation (\ref{eqa:part1}) by $\bD_j^{-1}$ gives
\begin{equation}\label{eqa:proof1}
-\bta_j \alpha_j +  \bD_j^{-1}\bJ\trasp_j\bS
\bS\bJ_j\bta_j\beta_j - \bD_j^{-1}\bR\trasp_j{\mathbf\bla} = \mathbf{0}
\end{equation}
Premultiplying the above by $\bR_j$ gives
\begin{equation*}
\bR_j\bD_j^{-1}\bJ\trasp_j\bS
\bS\bJ_j\bta_j\beta_j = \bR_j\bD_j^{-1}\bR\trasp_j{\mathbf\bla},
\end{equation*}
because of Equality (\ref{eqa:part2}). Therefore,
$$
\bla = (\bR_j\bD_j^{-1}\bR\trasp_j)^+ \bR_j\bD_j^{-1}\bJ\trasp_j\bS
\bS\bJ_j\bta_j\beta_j.
$$
Substituting this in Equation (\ref{eqa:proof1}) gives
\begin{equation*}
\bD_j^{-1}\bJ\trasp_j\bS
\bS\bJ_j\bta_j - \bD_j^{-1}\bR\trasp_j(\bR_j\bD_j^{-1}\bR\trasp_j)^+ \bR_j\bD_j^{-1}\bJ\trasp_j\bS
\bS\bJ_j\bta_j = \bta_j \frac{\alpha_j}{\beta_j}.
\end{equation*}
Since we want to maximize the function, the solution is the eigenvector corresponding to the maximum eigenvalue, which is equal to $\frac{\alpha_j}{\beta_j} =
\frac{\ba\trasp_j\bS\bS\ba_j}{\ba\trasp_j\bS\ba_j}$, as required, completing the proof.
\end{proof}
\subsubsection*{Correlated LS-SPCA components}
The approximate solutions are computed by requiring that each component gives the LS approximation of the residuals of $\bX$ orthogonal to the previously computed ones, $\bX_j = \bQ_{j}\bX $, with $\bX_1 = \bX$. Then, the loadings $\ba_j$ must satisfy:
\begin{align}\label{eqa:cspcals}
        &\ba_j = \argmin\limits_{{\ba}_j\, \in \Re^{p}}
        ||\bX_j - \bX\ba_j\bp\trasp_j|| =
\argmax\limits_{{{\ba}}_j \in \Re^{p}}
\frac{\ba\trasp_j\bS_j\bS_j\ba_j}{\ba\trasp_j\bS\ba_j}
 ,\; j = 1, \ldots, d,\\
&\text{subject to}\;
        a_{i,j} = 0\ {\rm if}\ i \not\in ind_j.\nonumber
\end{align}
The components following the first one do not maximise the variance explained
The correlated sparse components can be computed from the covariance matrix. If we let $\bZ_j = (\bI_p - \bA_{(j)}(\bA\trasp_{(j)}\bS\bA_{(j)})^{-1}\bA\trasp_{(j)}\bS)$, $\bZ_1=\bI_p$, the residual covariance matrix can be written as
%\begin{equation*}%\label{eqa:z}
$\bX_j\trasp\bX_j = \bS_j = \bS\bZ_j$. Then, the solutions are the eigenvectors satisfying:
\begin{equation}\label{eqa:cspcalssol}
\bD_j^{-1}\bJ\trasp_j\bZ\trasp_j\bS\bS\bZ_j\bJ_j\bta_j = \phi_{\text{max}} \bta_j.
\end{equation}
Obviously, the first component is the same as the uncorrelated one. The following solutions can be computed from the leftmost eigenvectors, $\bb_j$, of the matrices $\bD_j^{-1/2}\bJ_j\bZ\trasp_j\bS\bS\bZ_j\bJ_j\bD_j^{-1/2}$
as $\bta_j = \bD_j^{-1/2} \bb_j$.
\subsubsection*{Correlated LS-SPCA components with orthogonal loadings}
\begin{proposition}
The correlated sparse principal components of given indices $ind_j$ that minimize the $\Lt$ norm of the residuals of the approximation
subject to orthogonality constraints are defined as
\begin{align}\label{eqa:ospca}
        &\ba_j = \argmin\limits_{{\ba}_j\, \in \Re^{p}}
        ||\bX_j - \bX\ba_j\bp\trasp_j|| =
\argmax\limits_{{{\ba}}_j \in \Re^{p}}
\frac{\ba\trasp_j\bS_j\bS_j\ba_j}{\ba\trasp_j\bS\ba_j}
 ,\; j = 1, \ldots, d,\\
&{\rm subject\ to}\;
        a_{i,j} = 0\ {\rm if}\ i \not\in ind_j\;
         \rm{and}\; \ba\trasp_j\bA_j = \mathbf{0}\nonumber
\end{align}
The first solution is the same as for the uncorrelated case. The following ones are given by the eigenvectors satisfying
\begin{equation}\label{eqa:lsspcap}
    \bD_j^{-1}\bJ_j\trasp\left[\bI_{p} - \bA_j
    (\bA_j\trasp \bJ_j\bD_j^{-1}\bJ_j\trasp\bA_j)^{-1}
    \bA_j\trasp\bJ_j\bD_j^{-1}\bJ_j\trasp \right]
    \bS_j\bS_j\bJ_j\bta_j = \phi_{max}\bta_j,
\end{equation}
where $\phi_{\text{max}}$ is the largest eigenvalue.
\end{proposition}
\begin{proof}
Let $\bt_j = \bW_j\bta_j = \bX\bJ_j\bta_j$ be the $j$-th component. We need to maximize the variance explained subject to $\bA_j\trasp\bJ_j\bta_j = \mathbf{0}$. Hence we need to maximize the Langrangian:
\begin{align}
L(\bta_j, \bla) &= \text{tr}\left[\bX_j\trasp\bt(\bt\trasp_j\bt_j)^{-1}\bt\trasp_j\bX_j\right] - 2\bta_j\trasp\bJ\trasp_j\bA_j\bla\nonumber\\
&= \bta\trasp_j\bJ\trasp_j\bS_j \bS_j\bJ_j\bta_j(\bta\trasp_j\bD_j\bta_j)^{-1} - 2\bta_j\trasp\bJ_j\trasp\bA_j\bla
\end{align}
Equating the partial derivatives to zero gives:
\begin{align}
\frac{\partial L}{\partial \bta} &= -\bD_j\bta_j \alpha_j +  \bJ\trasp_j\bS_j
\bS_j\bJ_j\bta_j\beta_j - \bJ\trasp_j\bA_j\bla = \mathbf{0}\label{eqa:parto1}\\
\frac{\partial L}{\partial \bla} &= \bta_j\trasp\bA_j = \mathbf{0}\label{eqa:parto2}
\end{align}
where $\alpha_j = \bta\trasp_j\bJ\trasp_j\bS_j \bS_j\bJ_j\bta_j(\bta\trasp_j\bD_j\bta_j)^{-2}$ and $\beta_j = (\bta\trasp_j\bD_j\bta_j)^{-1}$. Premultiplying Equation (\ref{eqa:parto1}) by $\bD_j^{-1}$ gives
\begin{equation}\label{eqa:pro1}
\bta_j\alpha_j = \trasp\bD_j^{-1}\bJ\trasp_j\bS_j
\bS_j\bJ_j\bta_j\beta_j - \trasp\bD_j^{-1}\bA_j\bla
\end{equation}
Premultiplying the above by $\bA_j\trasp$ gives
$$
\bA_j\trasp\bD_j^{-1}\bJ\trasp_j\bS_j
\bS_j\bJ_j\bta_j\beta_j = \bA_j\trasp\bD_j^{-1}\bA_j\bla
$$
because of Equality (\ref{eqa:parto2}). Therefore,
$$
\bla = (\bA_j\trasp\bD_j^{-1}\bA_j)^{-1}\bA_j\trasp\bD_j^{-1}\bJ\trasp_j\bS_j
\bS_j\bJ_j\bta_j\beta_j.
$$
Substituting this into Equation (\ref{eqa:pro1}) gives
\begin{equation*}\label{eqa:lsspcap2}
    \bD_j^{-1}\bJ_j\trasp\left[\bI_{p} - \bA_j
    (\bA_j \bJ_j\bD_j^{-1}\bJ_j\trasp\bA_j)^{-1}
    \bA_j\trasp\bJ_j\bD_j^{-1}\bJ_j\trasp \right]
    \bS_j\bS_j\bJ_j\bta_j = \frac{\alpha_j}{\beta_j}\bta_j.
\end{equation*}
Since we want to maximize the function, the solution is the eigenvector corresponding to the maximum eigenvalue, which is equal to $\frac{\alpha_j}{\beta_j} =
\frac{\ba\trasp_j\bS_j\bS_j\ba_j}{\ba\trasp_j\bS\ba_j}$, as required in Equation (\ref{eqa:ospca}), completing the proof.
\end{proof} \end{document}